\documentclass[aps,prl,twocolumn,superscriptaddress,floatfix]{revtex4-2}
\usepackage[utf8]{inputenc}
\usepackage[english]{babel}
\usepackage{mathtools}
\usepackage{newtxtext}  
\usepackage{amsmath}
\usepackage{amsfonts}
\usepackage[varbb,varg]{newtxmath} 
\usepackage{bm,dsfont,eucal}
\AtBeginDocument{\renewcommand{\vec}[1]{\bm{#1}}}

\usepackage[dvipsnames,svgnames,x11names,hyperref]{xcolor}
\usepackage{hyperref}
\hypersetup{ 
    colorlinks=true, linkcolor=NavyBlue, urlcolor=NavyBlue, citecolor=NavyBlue
}
\usepackage{physics,braket,siunitx,slashed}
\usepackage[capitalise]{cleveref}

\begin{document}
        \title{Multicomponent Magnetic Domain Walls in Rhombohedral Graphene}

\author{Mainak Das}
\affiliation{Department of Physics and Astronomy, University of Kentucky, Lexington, Kentucky 40506-0055, USA}

\author{Nemin Wei}
\affiliation{Department of Physics and Yale Quantum Institute, Yale University, New Haven, Connecticut
06520, USA}

\author{Chunli Huang}
\affiliation{Department of Physics and Astronomy, University of Kentucky, Lexington, Kentucky 40506-0055, USA}


\begin{abstract}
Spatial textures of magnetic order, such as domain walls and skyrmions, are fundamental objects in magnetism. In rhombohedral multilayer graphene, magnetic order involves spin and valley degrees of freedom, opening the possibility of qualitatively new spatial textures. Here, we explore this possibility through a microscopic study of a one-dimensional domain wall in the valley-imbalanced quarter-metal phase of rhombohedral graphene. We uncover two different classes of domain walls. One resembles a conventional magnetic domain wall, locally rotating between the two bulk states, whereas the other is intrinsically multicomponent and explores states that are not occupied in either bulk domain. Which texture is realized is controlled by the competition between intervalley Hund's coupling and spin-orbit coupling, and we identify experimental signatures to distinguish them. 
For a superconducting junction formed across the wall, we further show that the superconducting phase difference couples directly to the intervalley-coherent phase of the texture, allowing precession of this internal phase to generate a voltage. This mechanism provides an explanation for the resistive domain walls recently observed in chiral superconductors [arXiv2605.13303] and predicts accompanying magnetization oscillations that could be detected using GHz-frequency AC probes. Our theory establish a microscopic connection between multicomponent magnetic textures, their internal dynamics, and superconducting transport.
\end{abstract}

\maketitle
\newpage
\textit{Introduction.--} In a conventional metallic ferromagnet (e.g.~nickel), the local order parameter is a three-dimensional spin magnetization, and a domain wall is described as a smooth rotation of this vector between two bulk orientations \cite{PhysRevLett.63.668,PhysRevB.43.3395,middelhoek1963domain,Yang_2022,cite-key1,cite-key2,alimohammadian2020observation}. In macroscopic magnetic materials, domain formation is determined by the competition between exchange, magnetic anisotropy, and long-range magnetostatic interactions. Dividing a uniform state into smaller domains reduces its magnetostatic energy at the cost of creating domain walls \cite{alma996296696802636,brown1965structure,alma9917490516802636,Venkat_2024}. Magnetism in rhombohedral graphene is fundamentally different in two respects. First, because the electron system is atomically thin, the conventional magnetostatic driving force for domain formation is strongly reduced. Second, the magnetization is not carried by spin alone: in the presence of an electric displacement field, spontaneous valley polarization also produces a substantial orbital magnetization \cite{geisenhof2021quantum,sheekey2026visualizingorbitalmagnetismelectron,han2023orbital,deng2026superconductivityferroelectricorbitalmagnetism,PhysRevB.107.L121405,PhysRevB.109.L060409}. 
At low carrier densities and low temperature, interactions can spontaneously polarize the spin and valley degrees of freedom, producing a quarter-metal state in which only one of the four spin-valley flavors is occupied     \cite{zhou2021half,arp2024intervalley,auerbach2025isospin}. Recent experiments have directly imaged magnetic domains in this regime \cite{sheekey2026visualizingorbitalmagnetismelectron,zhang2026imagingmeissnereffectlocal} and further suggest that the domain structure of the parent quarter metal is inherited by the chiral superconducting phase\cite{han2408signatures,dutta2026reconfigurable,hua2026multiknobswitchablechiralsuperconductivity,kalantre2026fermiologycandidatechiralsuperconductor}. Nevertheless, the microscopic structure of the domain walls remains largely unexplored. In particular, it is not known whether intrinsic spin-orbit coupling keeps spin and valley locked throughout the wall, as in transition-metal dichalcogenides \cite{sl5k-c825}, or whether they ``unlock'' inside the wall?

In this Letter, we address this question using a momentum-space micromagnetic calculation in rhombohedral graphene and a long-wavelength $CP^3$ theory. We uncover two qualitatively distinct classes of spin-valley domain walls, separated by a continuous transition driven by the competition between spin-orbit coupling and intervalley Hund's exchange, as shown in Fig.~\ref{2cdw_4cdw_occupation}. This transition occurs entirely within the domain wall while the adjoining bulk ferromagnetic states remain unchanged.
We identify experimental signatures that distinguish the two textures and further examine what becomes of these domain walls when chiral superconductivity develops in the adjoining domains. In this regime, the system forms an $S|\mathrm{IVC}|S$ junction, in which the intervalley-coherent domain-wall region acts as a magnetic weak link between two chiral superconductors. The resulting Josephson coupling directly links the superconducting phase difference to the internal intervalley-coherent phase of the magnetic texture.

\begin{figure}[t]
    \centering
    \includegraphics[width=\linewidth]{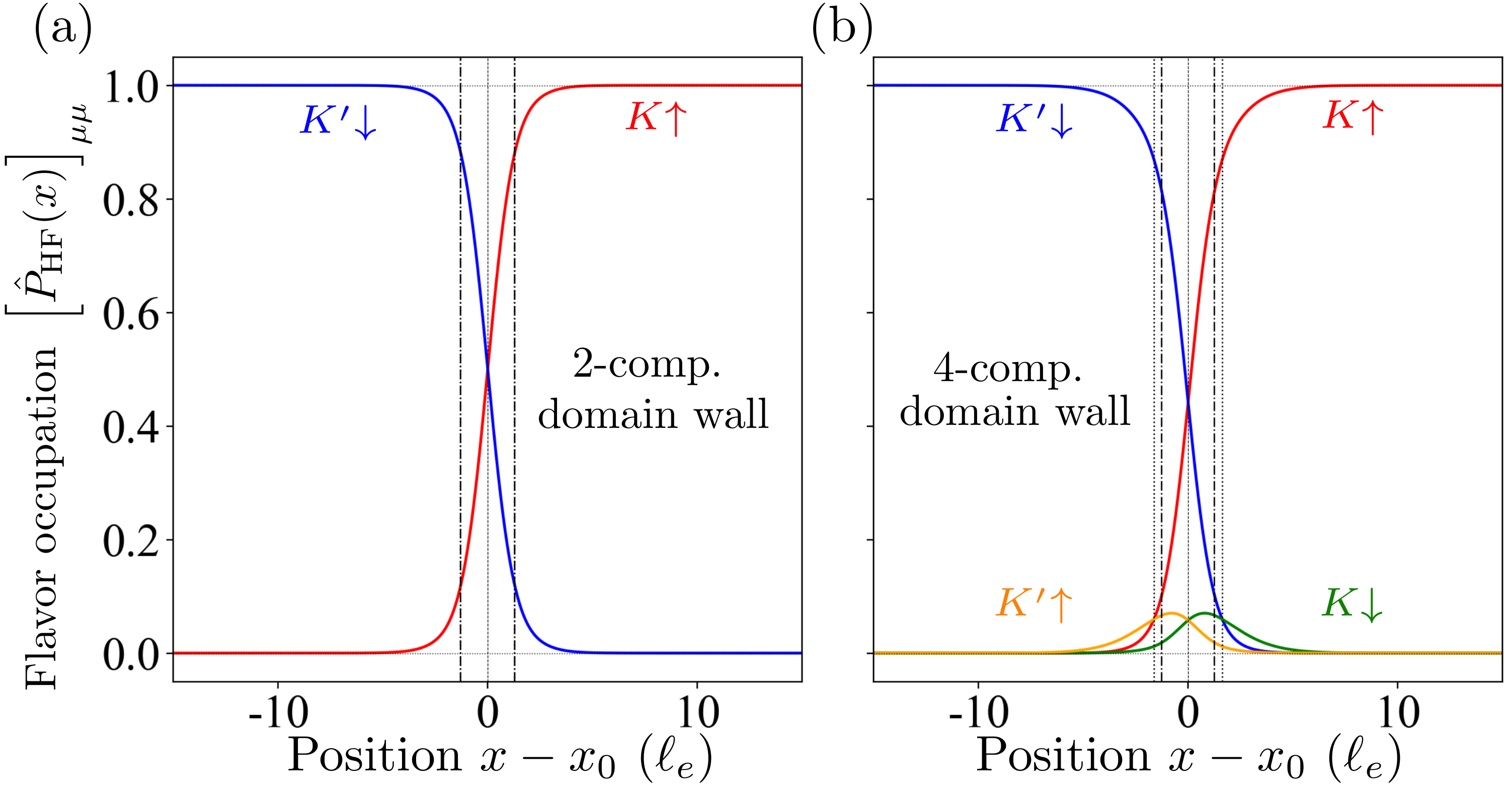}
\caption{
Two- and four-component domain walls in the valley-imbalanced quarter metal, determined by the competition between spin-orbit coupling $\lambda_{\rm soc}$ and intervalley Hund's coupling $g_\perp|n_e|$. In (a), $\lambda_{\rm soc}\gg g_\perp|n_e|$ the magnetic texture only has two flavors, whereas in (b), $\lambda_{\rm soc}\ll g_\perp|n_e|$ allows all four flavors to participate inside the wall. The curves show the diagonal component of the local spin-valley order parameter $[\hat P_{\rm HF}(x)]_{\mu\mu}$.}
    \label{2cdw_4cdw_occupation}
\end{figure}

\begin{figure*}[t]
    \centering
    \includegraphics[width=0.9\linewidth]{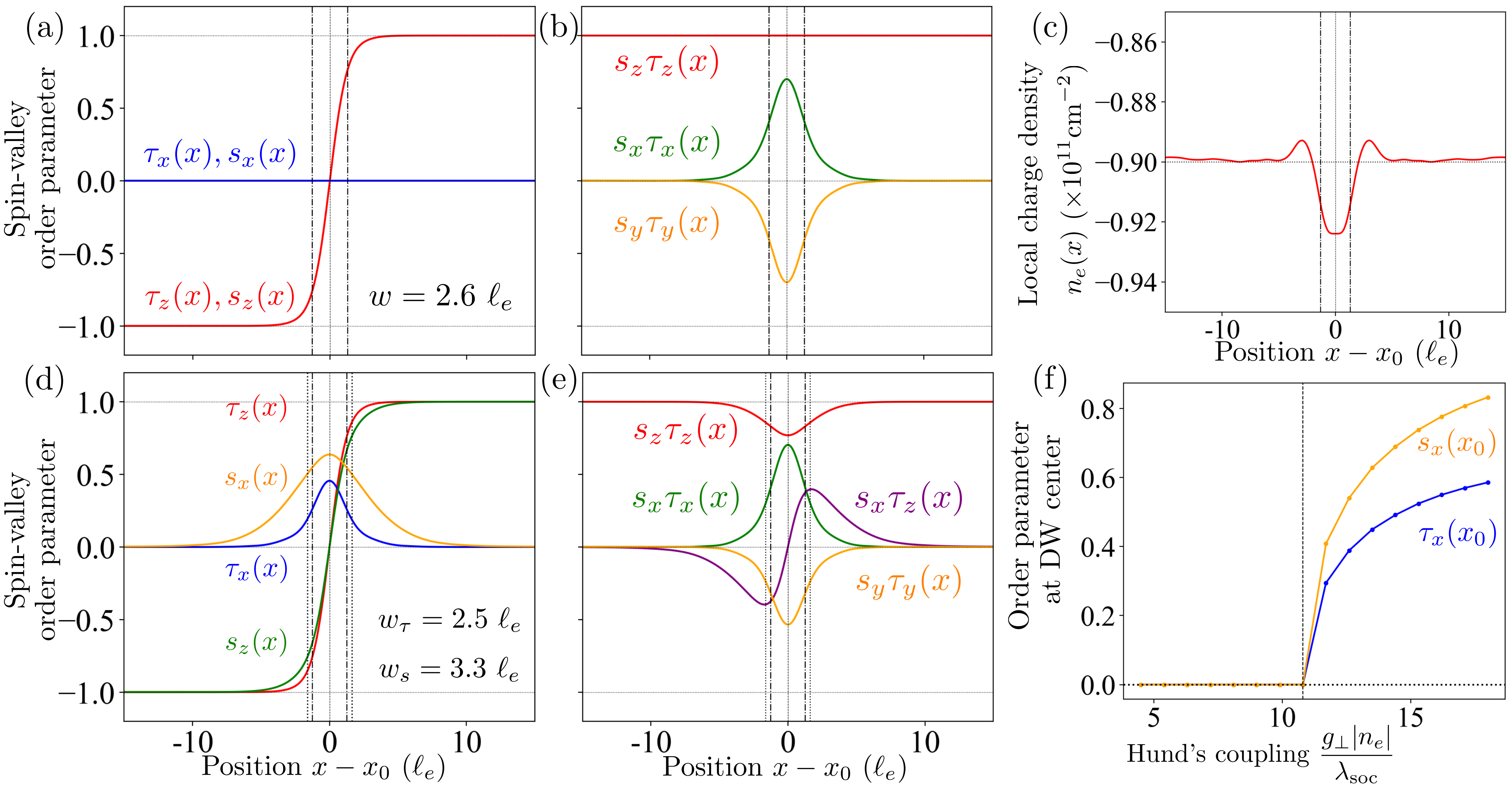}
    \caption{Self-consistent HF solutions of the two-component and the four-component spin--valley  domain walls.  (a–c) Results for the two-component domain wall at $g_\perp=0$ centered at $x=x_0$. (a,b) Local Spin–valley texture shows that the spin and valley
polarizations reverse simultaneously while remaining locked throughout the domain wall, giving rise to $s_z\tau_z(x)=1$ throughout the space. The characteristic length scale is the average hole spacing, $\ell_e=1/\sqrt{\pi|n_e|}\approx20~\rm{nm}$, and the domain-wall width $w=2.6\,\ell_e$ is indicated by the vertical dash-dotted lines. (c) Corresponding local charge density $n_e(x)$, showing only a weak redistribution of charge across the domain wall. The horizontal dashed line indicates the average charge density $n_e$. (d,e) Spin–valley texture of the four-component domain wall at $g_\perp|n_e|/\lambda_{\rm soc}=13.5$ shows  intervalley coherence $(\tau_x)$ and transverse spin polarization $(s_x)$ develop near the domain-wall center, partially releasing spin–valley locking $(s_z\tau_z<1)$ and producing distinct spin and valley domain-wall widths. The spin and valley widths are $w_s=3.3\,\ell_e$ (dotted lines) and $w_\tau=2.5\,\ell_e$(dash-dotted lines), respectively. (f) $s_x(x_0)$ and $\tau_x(x_0)$ at the domain wall center continuously grow beyond a critical coupling $g_\perp^*|n_e|/\lambda_{\rm soc}\approx\rm 11$.}
    \label{fig:cDWs}
\end{figure*}
 
\textit{Microscopic spin-valley domain walls.—}
Because the low-energy electrons carry four spin-valley flavors, their local magnetic order is described by a $4\times4$ Hermitian matrix. Excluding the total charge density, this matrix contains $4^2-1=15$ independent order parameters. Rather than parameterizing this large space and assuming a particular form  for the domain wall ~\cite{PhysRevLett.126.056801}, we determine the texture using unrestricted Hartree–Fock (HF) theory. This allows the charge density and all spin, valley, and mixed spin-valley order parameters to vary self-consistently in space.

We use rhombohedral bilayer graphene as a minimal model to identify magnetic domain-wall physics that we expect to persist in systems with other numbers of layers. We consider a one-dimensional domain wall for which the order parameter varies along $x$ while translational symmetry is preserved along $y$. To solve the problem numerically, periodic boundary conditions are imposed along the $x$ direction, such that the magnetic supercell forms a closed ring of circumference $L$. The periodic geometry therefore contains two equivalent domain walls separating the two degenerate bulk ferromagnetic states, with reciprocal vectors $G=2\pi m/L$. The in-plane coordinate axes are defined by taking $\vec e_x$ and $\vec e_y$ to be parallel and perpendicular, respectively, to the graphene Dirac-point momentum $\vec K$ measured from the $\Gamma$ point. In this basis, the self-consistent Hartree--Fock (HF) Hamiltonian is
\begin{align}
H^{\rm MF}_{G\alpha,G'\beta}(k_y)
={}&\delta_{GG'}
\left[
\hat H^{\rm SWMc}_{\alpha\beta}(
G\vec e_x+k_y \vec e_y)
+\hat H^{\rm SOC}_{\alpha\beta}
\right]
\nonumber\\
&+\Sigma^{\rm HF,C}_{G\alpha,G'\beta}(k_y)
+\Sigma^{\rm HF,Hund}_{G\alpha,G'\beta},
\label{eq:HMF_DW}
\end{align}
where $\alpha,\beta$ collectively label the spin $(s)$, valley $(\tau)$, layer $(l)$, and the A/B sublattice $(\sigma)$ degrees of freedom. The first two terms describe the microscopic SWMc band structure\cite{zhou2021half} and intrinsic spin-orbit coupling\cite{auerbach2025isospin}, while the last two are the HF self-energies arising from long-range Coulomb interactions and intervalley Hund's exchange. Details of the Hamiltonian and its numerical implementation are given in Ref.~\cite{supp_mat}. We first identify the density and displacement field for which the uniform HF ground state is a spin- and valley-polarized quarter metal\cite{PhysRevB.110.245118}. Intrinsic spin-orbit coupling locks the relative spin and valley orientations and reduces the fourfold flavor degeneracy to a time-reversal-related pair, which we denote by $K\uparrow$ and $K'\downarrow$. We then initialize the density matrix with a texture connecting these two states and solve Eq.~\eqref{eq:HMF_DW} self-consistently. The initial wall width parameter is estimated as $\xi\sim\sqrt{\rho_s/\mathcal{K}}$, where $\rho_s$ is the spin-valley stiffness and $\mathcal{K}$ is the magnetic anisotropy energy, both extracted from constrained uniform HF calculations \cite{supp_mat}. 

The converged eigenvalues and eigenvectors,
\begin{equation}
      H^{\rm MF}_{G\alpha,G'\beta}(k_y)z_{nG'\beta}(k_y)
    =\mathcal{E}_{n}(k_y)z_{nG\alpha}(k_y),
\end{equation}
determine the one-body density matrix $    \hat{\rho}(k_y)
    =
    \sum_n n_F(\mathcal{E}_{n}(k_y))
    |z_n(k_y)\rangle\langle z_{n}(k_y)|$.
To isolate the spin-valley order carried by the doped electrons, we define the doped-carrier density matrix $\hat {\tilde\rho}$ by restricting the band sum in $\hat \rho$ to the single-particle states of those doped carriers:
\begin{equation} 
\tilde\rho_{G\alpha,G'\beta}(k_y) = \sum_{n\in\mathcal D(k_y)} n_F\left( \mathcal E_n(k_y)\right) z_{nG\alpha}(k_y) z^*_{nG'\beta}(k_y).
\label{eq:doped_density_matrix} \end{equation} 
Here $\mathcal D(k_y)$ is the set of states filled as the chemical potential
is raised/lowered from charge neutrality to its physical (HF) value. 
Fourier transforming this density matrix along $x$ and tracing over the layer and sublattice degrees of freedom gives the normalized local spin-valley order parameter
\begin{align}
    [\hat P_{\rm HF}(x)]_{\mu\nu}
    ={}&
    \frac{1}{ n_{e}(x) L}
    \int\frac{dk_y}{(2\pi)}
    \sum_{G,G'}\sum_{a} \nonumber \\
    &\tilde{\rho}_{G\mu a,G'\nu a}(k_y)
    e^{i(G-G')x},
    \label{eq:su4_projector}
\end{align}
where $a\equiv(l,\sigma)$ runs through all the sublattices and $n_{e}(x)$ is the doped local charge density such that
$\Tr \hat P_{\rm HF}(x)=1$. Here,
$\mu,\nu\in\{K\uparrow,K'\uparrow,K\downarrow,K'\downarrow\}$. The diagonal elements of $\hat P$
give the local occupations of the four spin-valley flavors, as shown in Fig.~\ref{2cdw_4cdw_occupation}, and the local spin-valley order parameters are
\begin{equation}
     s_i\tau_j(x)
    =
    \Tr
    \left[\hat P_{\rm HF}(x)\hat s_i\hat \tau_j\right].
\end{equation}
This construction determines the complete spin-valley texture without assuming a particular rotation path across the domain wall.

The main results of our microscopic calculations are summarized in Fig.~\ref{fig:cDWs}. When spin-orbit coupling dominates the intervalley Hund's coupling, $\lambda_{\rm soc}\gg g_\perp|n_e|$, spin and valley remain locked throughout the wall. The texture is confined to the two-dimensional subspace favored by spin-orbit coupling, $K\uparrow$ and $K'\downarrow$. 
As shown in Figs.~\ref{fig:cDWs}(a) and \ref{fig:cDWs}(b), the spin-orbit locking condition $s_z \tau_z(x)=1$ remains satisfied at every position. We refer to this solution as the two-component domain wall. Although the two flavors form a superposition of opposite spin inside the domain wall, they belong to opposite valleys, and their coherence produces no net in-plane spin polarization. The wall therefore carries no static in-plane spin moment and has no direct linear Zeeman coupling to an in-plane magnetic field. This texture is adiabatically connected to the spin-valley-locked domain walls previously studied in transition-metal-dichalcogenides Chern insulators~\cite{sl5k-c825}. An important distinction is that the quarter-metal phase considered here is metallic and exhibits charge-density fluctuations across the wall, as shown in Fig.~\ref{fig:cDWs}(c).

The domain wall structure changes qualitatively in the opposite, Hund-dominated limit, $g_\perp|n_e|\gg\lambda_{\rm soc}$. Intervalley Hund's exchange favors parallel spin alignment in the two valleys and therefore admixes the other two flavors that are unoccupied in either bulk domains, $K\downarrow$ and $K'\uparrow$. The resulting spin-valley texture is shown in Figs.~\ref{fig:cDWs}(d) and (e). The reduction of $s_z\tau_z(x)$ below unity reflects a local spin-orbit-energy cost, which is compensated by the gain in intervalley exchange energy and accompanied by the emergence of transverse spin polarization $s_x(x)$ and intervalley coherence $\tau_x(x)$. Spin and valley are therefore no longer locked and develop distinct spatial profiles, with the spin-domain-wall width slightly exceeding the valley-domain-wall width. As shown in Fig.~\ref{fig:cDWs}(f), these additional order parameters turn on continuously above a critical Hund's coupling $g^*_\perp|n_e|/\lambda_{\rm soc}\sim11$, demonstrating a continuous transition from the two-component to the four-component domain wall.


As shown in companion experimental studies~\cite{josh2026preprint,josh2026preprint2}, the two textures exhibit distinct magnetotransport signatures: the 4-component domain wall carries a net in-plane spin moment and supports a ferro-Josephson effect whose critical current varies linearly with the applied in-plane magnetic field. 
Although our microscopic calculations focus on rhombohedral graphene, the same domain-wall physics may also arise in graphene moiré superlattices. This broader applicability motivates the universal long-wavelength theory outlined below and developed further in \cite{supp_mat}.

\begin{figure*}[t]
    \centering
    \includegraphics[width=\linewidth]{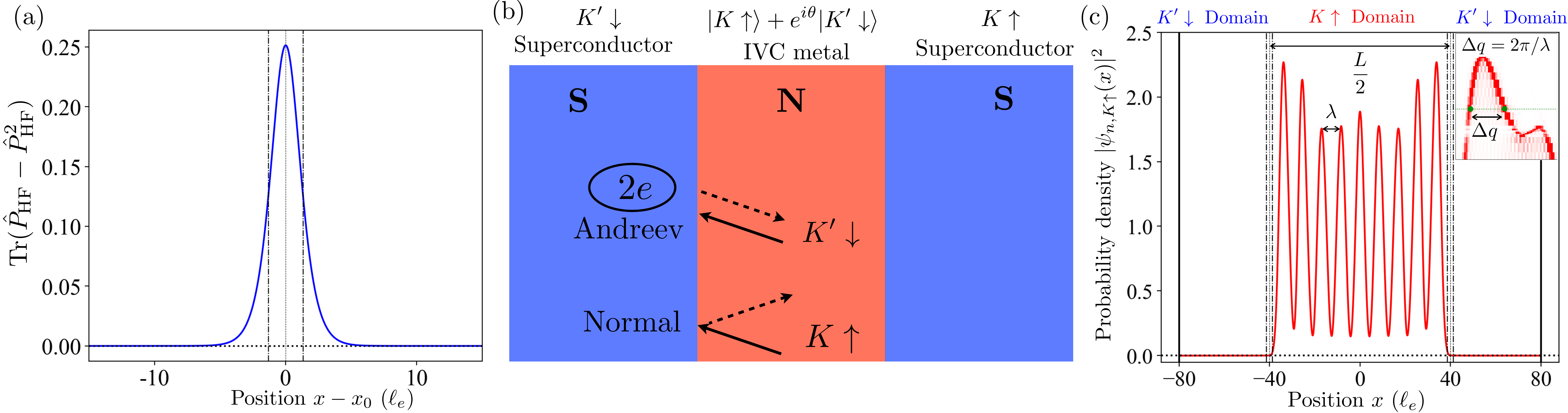}
    \caption{(a) Deviation of the microscopic HF spin--valley projector from the pure-state condition, quantified by $\text{Tr}(\hat P_{\rm HF}-\hat P_{\rm HF}^2)$. A finite value near the domain-wall center demonstrates that the local spin–valley density matrix is mixed, in contrast to the pure projector assumed in the $CP^3$ theory. (b) Schematic of spin–valley-selective scattering across an $S|N|S$ junction, where time-reversal-related chiral superconductors are separated by an intervalley-coherent(IVC) normal-metal, $\propto|K\uparrow\rangle+e^{i\theta}|K'\downarrow\rangle$. The IVC region coherently mixes the two valley flavors, enabling normal reflection and intervalley-assisted Andreev conversion.  (c) Probability density $|\psi_{n,K\uparrow}(x)|^2$ of a representative Bloch state at $\mathcal{E}_{n}(k_y=0)-\mu=5.6$ meV in the $K\uparrow$ domain. The bulk oscillation originates from the interference of two dominant Fourier components separated by momentum difference $\Delta q$, giving an oscillation wavelength $\lambda=2\pi/\Delta q$.}
    \label{fig:dw_fig_3}
\end{figure*}
\textit{$CP^3$ theory and its limitation.—}
A natural long-wavelength description, analogous to those used for
magic-angle twisted bilayer graphene~\cite{kwan2021domain} and quantum Hall
ferromagnets~\cite{lian20164}, represents the local spin-valley polarization by a
normalized four-component spinor $|\psi(x)\rangle$, or equivalently by the
rank-one projector
\begin{equation}
    \hat P(x)=|\psi(x)\rangle\langle\psi(x)|,
    \qquad
    \hat P^2=\hat P,\quad \Tr\hat P=1,
\end{equation}
whose order-parameter manifold is $CP^3$. We test this description using the
reduced spin-valley density matrix $\hat P_{\rm HF}(x)$ obtained from the
microscopic HF solution. As shown in
Fig.~\ref{fig:dw_fig_3}(a), $\hat P_{\rm HF}$ is nearly idempotent in the
bulk but becomes mixed inside the domain wall, as diagnosed by the finite
value of $\Tr(\hat P_{\rm HF}-\hat P_{\rm HF}^2)$. Microscopically, this arises because the intervalley coherence and layer-sublattice degrees of
freedom become entangled inside the domain wall. The $CP^3$ theory therefore
captures the qualitative spin-valley texture but not the full microscopic
quasiparticle wavefunctions. For example, the idealized $CP^3$ texture would have predicted
$|s_x\tau_x(x_0)|=|s_y\tau_y(x_0)|=1$ at the center of two-component domain wall (Fig.~\ref{fig:cp3_DWs} in Ref.\cite{supp_mat}),  whereas
their magnitudes are reduced in the microscopic HF solution (Fig.~\ref{fig:cDWs}b).

\textit{Chiral superconductivity across domain walls:--} One of the most striking aspects of chiral superconductivity in the quarter-metal regime is the apparent intertwining of magnetism and superconductivity \cite{han2025signatures,hua2026multi,dutta2026reconfigurable,sheekey2026visualizing}. In Ref.~\cite{han2025signatures}, it was found that the nominally zero-resistance state shows resistance spikes $(R_{xx}\ll h/e^2) $ near the coercive field associated with magnetization reversal of the parent quarter metal. NanoSQUID imaging further reveals that the chiral domain structure of the superconducting phase is inherited from its valley-polarized parent metal and that the domain walls are resistive~\cite{dutta2026reconfigurable}.  These observations suggest that, below $T_c$, a normal-state spin-valley domain wall evolves into a weak link between superconducting domains of opposite chirality.

This picture raises two basic questions. First, what produces the observed resistance? A conventional $S|N|S$ junction can carry a dissipationless Josephson current mediated by Andreev bound states, even when the normal region has finite resistance. The presence of a normal domain-wall region is therefore not, by itself, sufficient to produce a voltage across the junction. Moreover, because the domain wall is only a few interparticle spacings wide, the junction can, in principle, support a supercurrent below its critical current. 

Second, when Cooper pairs tunnel across the domain wall, what microscopic process converts pairs with center-of-mass momentum $2K$ into pairs with momentum $2K'$\cite{Qin_2026}? Although the domain wall breaks translational symmetry, it varies over the interparticle spacing, which is much larger than the atomic scale, and therefore cannot by itself provide the large momentum transfer required for intervalley conversion. We argue below that these two question have a common resolution: the superconducting phase couples dynamically to the intervalley-coherent phase within the domain wall.

Let's consider the superconducting junction shown in Fig.~\ref{fig:dw_fig_3}(b). The left and right regions are time-reversal-related chiral superconductors with order parameters $\Delta_L = |\Delta_L|e^{i\phi_L}$, $\Delta_R = |\Delta_R|e^{i\phi_R}$ formed respectively from electrons in the $K$ and $K'$ valleys. We model the domain-wall as a non-superconducting intervalley-coherent state with order parameter $
    M \equiv \langle c_K^\dagger c_{K'}\rangle
    =|M|e^{i\theta}$
where $\theta$ is the intervalley-coherent phase.  To an excellent approximation, the low-energy Hamiltonian separately conserves the electron numbers in the two valleys. 
Because the microscopic Hamiltonian is invariant under independent valley rotations, $ c_{\tau}\rightarrow e^{i\alpha_\tau}c_\tau$ ($\tau=K,K'$),
the phases of the three order parameters transform as
\begin{equation}
(\phi_L,\phi_R,\theta)
\rightarrow
(\phi_L+2\alpha_K,\,
 \phi_R+2\alpha_{K'},\,
 \theta+\alpha_{K'}-\alpha_K).
\end{equation}
Consequently, the Josephson energy can  depend on the phases only through the valley-invariant combination 
\begin{equation}
    \Phi=\varphi-2\theta\;\;,\;\;\varphi=\phi_R-\phi_L
    -\frac{2e}{\hbar}\int_L^R\mathbf{A}\cdot d\boldsymbol{\ell}
\end{equation}
To leading harmonic order, the phase-dependent free energy then takes the form
\begin{equation}
    F=F_J+F_U=-E_J\cos(\varphi-2\theta)
    - E_{U}\cos(3\theta),
\end{equation}
where $E_J$ describes Josephson coupling energy across the domain wall and 
$E_U$ is the pinning potential generated by 3-body Umklapp processes, whose strength is estimated in Ref.~\cite{das2025momentum}.
The corresponding Josephson current is
\begin{equation}
    I_s=\frac{2e}{\hbar}\frac{\partial F}{\partial\varphi}
       =I_c\sin\Phi,
    \qquad
    I_c=\frac{2eE_J}{\hbar}.
\end{equation}
Because $\theta$ enters the same invariant phase combination, Josephson coupling exerts a torque on the
intervalley-coherent phase,
\begin{equation}
    \tau_\theta=-\frac{\partial F}{\partial\theta}
    =\frac{\hbar}{e}I_s-3E_U\sin(3\theta).
\end{equation}
The voltage drop across the junction is related to the dynamics of both phase
variables through the Josephson relation,
\begin{equation}
    V=\frac{\hbar}{2e}\dot{\varphi}
      =\frac{\hbar}{2e}\left(\dot{\Phi}+2\dot{\theta}\right).
\end{equation}
This equation implies there are two distinct critical currents. The first is the
usual Josephson critical current, $    I_c=\frac{2eE_J}{\hbar},$
above which 
$\Phi$ starts precessing. The second is the depinning current of the intervalley-coherent phase, $ I_c'
    =\frac{3eE_U}{\hbar}$, above which $\theta$ starts precessing.
A fully stationary state therefore exists only when the applied current $I<\min(I_c,I_c')$.
Thus, for the $S|IVC|S$ junction, even below the conventional Josephson critical current, the junction can becomes resistive once
IVC order parameter is non-stationary.
Our analysis therefore suggests that the resistance observed in
Ref.~\cite{dutta2026reconfigurable} may originate either from conventional
Josephson breakdown or from current-driven dynamics of the
intervalley-coherent phase, depending on which critical current is smaller.
A possible test is to measure the nonlinear $IV$ characteristic: below the depinning threshold, Andreev bound states carry a dissipationless supercurrent; above threshold, phase precession generates a finite voltage. Importantly, IVC precession also produces GHz oscillations of orbital magnetization which can be  detected using AC probes \cite{das2025momentum}.

\textit{Interference pattern:--} Beyond their collective dynamics, the metallic domain walls also produce interference pattern for the stationary quasiparticle states. As shown in
Fig.~\ref{fig:dw_fig_3}(c), the probability density
$|\psi_{K\uparrow}(x)|^2$ of a stationary state in a $K\uparrow$ domain bounded
by two $K'\downarrow$ domains shows pronounced spatial oscillations. These
oscillations arise from interference between the two equal-energy momenta
$k_{x,1}$ and $k_{x,2}$ shown in the inset. Their separation,
$\Delta q=|k_{x,1}-k_{x,2}|$, determines the interference wavelength: $\lambda=\frac{2\pi}{\Delta q}$ and $ N\lambda\sim\frac{L}{2}$. Here $L/2$ is the length of the $K\uparrow$ domain and $N$ is the number of
oscillation periods across it. For the state shown in
Fig.~\ref{fig:dw_fig_3}(c), $N=9$. This energy-dependent interference pattern
provides a microscopic signature of metallic domain structure that could be
detected using spatially resolved probes.

\textit{Summary and outlook:-} We have developed a micromagnetic theory in momentum-space to study magnetic textures in rhombohedral graphene and uncovered two qualitatively distinct multicomponent spin-valley domain walls. Their competition is determined by the competition between spin-orbit coupling and intervalley Hund's exchange and occurs entirely within the wall while the adjoining bulk states remain unchanged.  This work provides a starting point for studying more general spin-valley textures \cite{cpl_43_2_020714}, including skyrmions, as well as interactions between domain walls and quasiparticles.

\textit{Note added.—}
During the preparation of this manuscript, we became aware of a related
work Ref.~\cite{phong2026valley} on domain walls in rhombohedral multilayer
graphene. That work studies transport across a prescribed valley-domain-wall
texture, while we determine the domain-wall structure microscopically. Ref.~\cite{archisman2026preprint} develops a theory of hydrodynamic transport across orbital-ferromagnetic domain walls. Refs.~\cite{josh2026preprint,josh2026preprint2} report experimental evidence of unconventional transport across such domain walls.

\textit{Acknowledgments.—} We thank Ruiheng Su and Joshua Folk for insightful discussions and for sharing experimental data that inspired this theoretical work. M.D. and C.H. were supported in part by the U.S. Department of Energy, Office of Science, Office of Basic Energy Sciences, under Award No.~DE-SC0024346 and in part by the National Science Foundation CAREER Award No.~DMR-2541471.
We acknowledge the University of Kentucky Center for Computational Sciences and Information Technology Services Research Computing for their support and use of the
Morgan Compute Cluster and associated research computing resources. 

\bibliographystyle{ieeetr}

\bibliography{references}

\begin{thebibliography}{10}

\bibitem{PhysRevLett.63.668}
M.~R. Scheinfein, J.~Unguris, R.~J. Celotta, and D.~T. Pierce, ``Influence of
  the surface on magnetic domain-wall microstructure,'' {\em Phys. Rev. Lett.},
  vol.~63, pp.~668--671, Aug 1989.

\bibitem{PhysRevB.43.3395}
M.~R. Scheinfein, J.~Unguris, J.~L. Blue, K.~J. Coakley, D.~T. Pierce, R.~J.
  Celotta, and P.~J. Ryan, ``Micromagnetics of domain walls at surfaces,'' {\em
  Phys. Rev. B}, vol.~43, pp.~3395--3422, Feb 1991.

\bibitem{middelhoek1963domain}
S.~Middelhoek, ``Domain walls in thin ni--fe films,'' {\em Journal of Applied
  Physics}, vol.~34, no.~4, pp.~1054--1059, 1963.

\bibitem{Yang_2022}
H.-H. Yang, N.~Bansal, P.~Rüßmann, M.~Hoffmann, L.~Zhang, D.~Go, Q.~Li, A.-A.
  Haghighirad, K.~Sen, S.~Blügel, M.~Le~Tacon, Y.~Mokrousov, and W.~Wulfhekel,
  ``Magnetic domain walls of the van der waals material fe3gete2,'' {\em 2D
  Materials}, vol.~9, p.~025022, mar 2022.

\bibitem{cite-key1}
B.~Huang, G.~Clark, E.~Navarro-Moratalla, D.~R. Klein, R.~Cheng, K.~L. Seyler,
  D.~Zhong, E.~Schmidgall, M.~A. McGuire, D.~H. Cobden, W.~Yao, D.~Xiao,
  P.~Jarillo-Herrero, and X.~Xu, ``Layer-dependent ferromagnetism in a van der
  waals crystal down to the monolayer limit,'' {\em Nature}, vol.~546,
  no.~7657, pp.~270--273, 2017.

\bibitem{cite-key2}
C.~Gong, L.~Li, Z.~Li, H.~Ji, A.~Stern, Y.~Xia, T.~Cao, W.~Bao, C.~Wang,
  Y.~Wang, Z.~Q. Qiu, R.~J. Cava, S.~G. Louie, J.~Xia, and X.~Zhang,
  ``Discovery of intrinsic ferromagnetism in two-dimensional van der waals
  crystals,'' {\em Nature}, vol.~546, no.~7657, pp.~265--269, 2017.

\bibitem{alimohammadian2020observation}
M.~Alimohammadian and B.~Sohrabi, ``Observation of magnetic domains in graphene
  magnetized by controlling temperature, strain and magnetic field,'' {\em
  Scientific Reports}, vol.~10, no.~1, p.~21325, 2020.

\bibitem{alma996296696802636}
W.~F. Brown, {\em Micromagnetics.}
\newblock Interscience tracts on physics and astronomy, no. 18, New York:
  Interscience Publishers, 1963.

\bibitem{brown1965structure}
W.~F. Brown~Jr and A.~E. LaBonte, ``Structure and energy of one-dimensional
  domain walls in ferromagnetic thin films,'' {\em Journal of Applied Physics},
  vol.~36, no.~4, pp.~1380--1386, 1965.

\bibitem{alma9917490516802636}
A.~Hubert and R.~Schäfer, {\em Magnetic domains : the analysis of magnetic
  microstructures}.
\newblock Berlin ;: Springer, 1998.

\bibitem{Venkat_2024}
G.~Venkat, D.~A. Allwood, and T.~J. Hayward, ``Magnetic domain walls: types,
  processes and applications,'' {\em Journal of Physics D: Applied Physics},
  vol.~57, p.~063001, nov 2023.

\bibitem{geisenhof2021quantum}
F.~R. Geisenhof, F.~Winterer, A.~M. Seiler, J.~Lenz, T.~Xu, F.~Zhang, and R.~T.
  Weitz, ``Quantum anomalous hall octet driven by orbital magnetism in bilayer
  graphene,'' {\em Nature}, vol.~598, no.~7879, pp.~53--58, 2021.

\bibitem{sheekey2026visualizingorbitalmagnetismelectron}
O.~I. Sheekey, T.~B. Arp, B.~A. Foutty, R.~Zhang, T.~Tan, L.~F.~W. Holleis,
  Y.~Guo, S.~S. Kalantre, C.~Zhang, M.~Zakharyan, D.~Gong, A.~Keough, Y.~Choi,
  Y.~Choi, S.~Xu, T.~Xie, B.~H. Alexander, M.~Hocking, Q.~Cao, M.~E. Huber,
  T.~Taniguchi, K.~Watanabe, C.~Jin, E.~Lantagne-Hurtubise, A.~Sharpe,
  T.~Devakul, and A.~F. Young, ``Visualizing orbital magnetism in electron
  doped rhombohedral multilayer graphene,'' 2026.

\bibitem{han2023orbital}
T.~Han, Z.~Lu, G.~Scuri, J.~Sung, J.~Wang, T.~Han, K.~Watanabe, T.~Taniguchi,
  L.~Fu, H.~Park, and L.~Ju, ``Orbital multiferroicity in pentalayer
  rhombohedral graphene,'' {\em Nature}, vol.~623, no.~7985, pp.~41--47, 2023.

\bibitem{deng2026superconductivityferroelectricorbitalmagnetism}
J.~Deng, J.~Xie, H.~Li, T.~Taniguchi, K.~Watanabe, J.~Shan, K.~F. Mak, and
  X.~Liu, ``Superconductivity and ferroelectric orbital magnetism in
  semimetallic rhombohedral hexalayer graphene,'' 2026.

\bibitem{PhysRevB.107.L121405}
C.~Huang, T.~M.~R. Wolf, W.~Qin, N.~Wei, I.~V. Blinov, and A.~H. MacDonald,
  ``Spin and orbital metallic magnetism in rhombohedral trilayer graphene,''
  {\em Phys. Rev. B}, vol.~107, p.~L121405, Mar 2023.

\bibitem{PhysRevB.109.L060409}
M.~Das and C.~Huang, ``Unconventional metallic ferromagnetism: Nonanalyticity
  and sign-changing behavior of orbital magnetization in rhombohedral trilayer
  graphene,'' {\em Phys. Rev. B}, vol.~109, p.~L060409, Feb 2024.

\bibitem{zhou2021half}
H.~Zhou, T.~Xie, A.~Ghazaryan, T.~Holder, J.~R. Ehrets, E.~M. Spanton,
  T.~Taniguchi, K.~Watanabe, E.~Berg, M.~Serbyn, {\em et~al.}, ``Half-and
  quarter-metals in rhombohedral trilayer graphene,'' {\em Nature}, vol.~598,
  no.~7881, pp.~429--433, 2021.

\bibitem{arp2024intervalley}
T.~Arp, O.~Sheekey, H.~Zhou, C.~Tschirhart, C.~L. Patterson, H.~Yoo,
  L.~Holleis, E.~Redekop, G.~Babikyan, T.~Xie, {\em et~al.}, ``Intervalley
  coherence and intrinsic spin--orbit coupling in rhombohedral trilayer
  graphene,'' {\em Nature Physics}, vol.~20, no.~9, pp.~1413--1420, 2024.

\bibitem{auerbach2025isospin}
N.~Auerbach, S.~Dutta, M.~Uzan, Y.~Vituri, Y.~Zhou, A.~Y. Meltzer, S.~Grover,
  T.~Holder, P.~Emanuel, M.~E. Huber, Y.~Myasoedov, K.~Watanabe, T.~Taniguchi,
  Y.~Oreg, E.~Berg, and E.~Zeldov, ``Isospin magnetic texture and intervalley
  exchange interaction in rhombohedral tetralayer graphene,'' {\em Nature
  Physics}, vol.~21, no.~11, pp.~1765--1772, 2025.

\bibitem{zhang2026imagingmeissnereffectlocal}
R.~Zhang, B.~A. Foutty, O.~Sheekey, T.~Arp, S.~Xu, T.~Xie, Y.~Guo, H.~Stoyanov,
  S.~Gu, A.~Keough, E.~Redekop, C.~Zhang, T.~Taniguchi, K.~Watanabe, M.~E.
  Huber, C.~Jin, E.~Berg, and A.~F. Young, ``Imaging the meissner effect and
  local superfluid stiffness in a graphene superconductor,'' 2026.

\bibitem{han2408signatures}
T.~Han, Z.~Lu, Z.~Hadjri, L.~Shi, Z.~Wu, W.~Xu, Y.~Yao, A.~A. Cotten,
  O.~Sharifi~Sedeh, H.~Weldeyesus, J.~Yang, J.~Seo, S.~Ye, M.~Zhou, H.~Liu,
  G.~Shi, Z.~Hua, K.~Watanabe, T.~Taniguchi, P.~Xiong, D.~M. Zumb{\"u}hl,
  L.~Fu, and L.~Ju, ``Signatures of chiral superconductivity in rhombohedral
  graphene,'' {\em Nature}, vol.~643, no.~8072, pp.~654--661, 2025.

\bibitem{dutta2026reconfigurable}
S.~Dutta, N.~Auerbach, T.~Han, Y.~Zhou, G.~Shavit, N.-S. Kander, Y.~Myasoedov,
  M.~E. Huber, K.~Watanabe, T.~Taniguchi, {\em et~al.}, ``Reconfigurable chiral
  superconductivity,'' {\em arXiv preprint arXiv:2605.13303}, 2026.

\bibitem{hua2026multiknobswitchablechiralsuperconductivity}
Z.~Hua, S.~Ye, P.~Pattanakanvijit, G.~Shi, T.~Han, E.~Aitken, J.~Yang, J.~Seo,
  H.~Liu, R.~Hao, K.~Xiao, J.~Guo, V.~T. Phong, K.~Watanabe, T.~Taniguchi,
  C.~Huang, C.~Lewandowski, L.~Ju, P.~Xiong, and Z.~Lu, ``Multi-knob switchable
  chiral superconductivity quartet in rhombohedral graphene,'' 2026.

\bibitem{kalantre2026fermiologycandidatechiralsuperconductor}
S.~S. Kalantre, B.~H. Alexander, J.~May-Mann, J.~Herzog-Arbeitman, M.~Hocking,
  Q.~Cao, K.~Watanabe, T.~Taniguchi, D.~Goldhaber-Gordon, A.~J. Mannix,
  T.~Devakul, Y.~H. Kwan, D.~E. Parker, and A.~Sharpe, ``Fermiology and the
  candidate chiral superconductor in rhombohedral tetralayer graphene,'' 2026.

\bibitem{sl5k-c825}
W.-X. Qiu and F.~Wu, ``Topological magnons and domain walls in twisted bilayer
  ${\mathrm{mote}}_{2}$,'' {\em Phys. Rev. B}, vol.~112, p.~085132, Aug 2025.

\bibitem{PhysRevLett.126.056801}
C.~Huang, N.~Wei, and A.~H. MacDonald, ``Current-driven magnetization reversal
  in orbital chern insulators,'' {\em Phys. Rev. Lett.}, vol.~126, p.~056801,
  Feb 2021.

\bibitem{supp_mat}
See Supplemental Material for details of mean-field construction of uniform and
  domain wall solutions, and reconstruction of domain wall profiles using
  $CP^3$ theory .

\bibitem{PhysRevB.110.245118}
J.~M. Koh, A.~Thomson, J.~Alicea, and E.~Lantagne-Hurtubise, ``Symmetry-broken
  metallic orders in spin-orbit-coupled bernal bilayer graphene,'' {\em Phys.
  Rev. B}, vol.~110, p.~245118, Dec 2024.

\bibitem{josh2026preprint}
R.~Su, Z.~Gao, C.~Coleman, M.~Kuiri, D.~Waters, K.~Watanabe, T.~Taniguchi,
  M.~Yankowitz, N.~Wei, C.~Huang, A.~H. MacDonald, and J.~Folk, ``Signatures of
  a ferro-josephson effect in twisted graphene,'' {\em arXiv:2608.25257}, 2026.

\bibitem{josh2026preprint2}
Z.~Gao, C.~Coleman, S.~Folk, R.~Su, M.~Kuiri, K.~Watanabe, T.~Taniguchi,
  N.~Wei, C.~Huang, and J.~Folk, ``Metastable magnetic domains and the
  anomalous $b_\parallel=0$ resistance peak in twisted double bilayer
  graphene,'' {\em arXiv:2608.25263}, 2026.

\bibitem{kwan2021domain}
Y.~H. Kwan, G.~Wagner, N.~Chakraborty, S.~H. Simon, and S.~Parameswaran,
  ``Domain wall competition in the chern insulating regime of twisted bilayer
  graphene,'' {\em Physical Review B}, vol.~104, no.~11, p.~115404, 2021.

\bibitem{lian20164}
Y.~Lian, A.~Rosch, and M.~Goerbig, ``Su(4) skyrmions in the $\nu=\pm1$ quantum
  hall state of graphene,'' {\em Physical Review Letters}, vol.~117, no.~5,
  p.~056806, 2016.

\bibitem{han2025signatures}
T.~Han, Z.~Lu, Z.~Hadjri, L.~Shi, Z.~Wu, W.~Xu, Y.~Yao, A.~A. Cotten,
  O.~Sharifi~Sedeh, H.~Weldeyesus, {\em et~al.}, ``Signatures of chiral
  superconductivity in rhombohedral graphene,'' {\em Nature}, vol.~643,
  no.~8072, pp.~654--661, 2025.

\bibitem{hua2026multi}
Z.~Hua, S.~Ye, P.~Pattanakanvijit, G.~Shi, T.~Han, E.~Aitken, J.~Yang, J.~Seo,
  H.~Liu, R.~Hao, {\em et~al.}, ``Multi-knob switchable chiral
  superconductivity quartet in rhombohedral graphene,'' {\em arXiv preprint
  arXiv:2607.06520}, 2026.

\bibitem{sheekey2026visualizing}
O.~I. Sheekey, T.~B. Arp, B.~A. Foutty, R.~Zhang, T.~Tan, L.~F. Holleis,
  Y.~Guo, S.~S. Kalantre, C.~Zhang, M.~Zakharyan, {\em et~al.}, ``Visualizing
  orbital magnetism in electron doped rhombohedral multilayer graphene,'' {\em
  arXiv preprint arXiv:2605.30316}, 2026.

\bibitem{Qin_2026}
Q.~Qin and C.~Wu, ``Chiral finite-momentum superconductivity in the tetralayer
  graphene,'' {\em Chinese Physics Letters}, vol.~43, p.~030708, mar 2026.

\bibitem{das2025momentum}
M.~Das and C.~Huang, ``Momentum space ac josephson effect and intervalley
  coherence in multilayer graphene,'' {\em Nature Communications}, vol.~17,
  no.~1, p.~1079, 2025.

\bibitem{cpl_43_2_020714}
Y.~Zhu, Z.~Li, Y.~Xu, Y.~Ling, Z.~Lin, Z.~Chen, Z.~Liu, H.~Fu, and X.~Lin,
  ``Charge excitations at landau level crossings in
  wse<sub>2</sub>-encapsulated bilayer graphene,'' {\em Chin. Phys. Lett.},
  vol.~43, no.~2, pp.~020714--020714, 2026.

\bibitem{phong2026valley}
V.~T. Phong, E.~Prada, P.~San-Jose, F.~Guinea, and E.~J. Mele, ``Valley valves
  at domain walls in symmetry-broken rhombohedral graphene,'' {\em arXiv
  preprint arXiv:2606.14878}, 2026.

\bibitem{archisman2026preprint}
A.~Panigrahi and K.~Nazaryan, ``Viscochiral transport: Chiral selection of
  hydrodynamic vortices by berry curvature,'' {\em arXiv:2608.26102}, 2026.

\bibitem{PhysRevB.111.125127}
G.~Xu and C.~Huang, ``Influence of the dirac sea on phase transitions in
  monolayer graphene under strong magnetic fields,'' {\em Phys. Rev. B},
  vol.~111, p.~125127, Mar 2025.

\bibitem{de2023global}
S.~J. De, A.~Das, S.~Rao, R.~K. Kaul, and G.~Murthy, ``Global phase diagram of
  charge-neutral graphene in the quantum hall regime for generic
  interactions,'' {\em Physical Review B}, vol.~107, no.~12, p.~125422, 2023.

\bibitem{aleiner2007spontaneous}
I.~Aleiner, D.~Kharzeev, and A.~Tsvelik, ``Spontaneous symmetry breaking in
  graphene subjected to an in-plane magnetic field,'' {\em Physical Review
  B—Condensed Matter and Materials Physics}, vol.~76, no.~19, p.~195415,
  2007.

\bibitem{lemonik2010spontaneous}
Y.~Lemonik, I.~Aleiner, and V.~Fal’Ko, ``Spontaneous symmetry breaking and
  lifshitz transition in bilayer graphene,'' {\em Physical Review B—Condensed
  Matter and Materials Physics}, vol.~82, no.~20, p.~201408, 2010.

\bibitem{wei2025landau}
N.~Wei, G.~Xu, I.~S. Villadiego, and C.~Huang, ``Landau-level mixing and su (4)
  symmetry breaking in graphene,'' {\em Physical Review Letters}, vol.~134,
  no.~4, p.~046501, 2025.

\bibitem{kharitonov2012phase}
M.~Kharitonov, ``Phase diagram for the $\nu$= 0 quantum hall state in monolayer
  graphene,'' {\em Physical Review B—Condensed Matter and Materials Physics},
  vol.~85, no.~15, p.~155439, 2012.

\bibitem{chatterjee2022inter}
S.~Chatterjee, T.~Wang, E.~Berg, and M.~P. Zaletel, ``Inter-valley coherent
  order and isospin fluctuation mediated superconductivity in rhombohedral
  trilayer graphene,'' {\em Nature Communications}, vol.~13, no.~1, p.~6013,
  2022.

\bibitem{kolavr2026electrostatically}
K.~Kol{\'a}{\v{r}}, A.~F. Young, and C.~Lewandowski, ``Electrostatically
  stabilized surface flat bands in rhombohedral graphite at zero displacement
  field,'' {\em arXiv preprint arXiv:2605.24080}, 2026.

\bibitem{wolf2024magnetism}
T.~Wolf, N.~Wei, H.~Zhou, and C.~Huang, ``Magnetism in the dilute electron gas
  of rhombohedral multilayer graphene,'' {\em arXiv preprint arXiv:2408.15884},
  2024.

\end{thebibliography}
\nocite{PhysRevB.111.125127}
\newpage

\setcounter{equation}{0}
\setcounter{figure}{0}
\setcounter{table}{0}
\setcounter{page}{1}
\makeatletter
\renewcommand{\theequation}{S\arabic{equation}}
\renewcommand{\thefigure}{S\arabic{figure}}

\maketitle
\widetext
\begin{center}
\textbf{\large Supplementary Materials: Multicomponent Magnetic Domain Walls in Rhombohedral Graphene}
\end{center}
This Supplementary Material is organized as follows. In Sec.~A, we present the microscopic Hartree–Fock (HF) theory of the uniform quarter-metal ground state, including the phase diagram and the transition between the two- and four-component intervalley-coherent (IVC) phases. Sec.~B describes the HF formulation for spin–valley domain walls in a one-dimensional magnetic supercell, together with the numerical convergence and microscopic characterization of the domain-wall solutions. Section~C presents the long-wavelength $CP^3$ theory and compares its predictions with the microscopic HF results.

We first summarize the microscopic continuum model and its uniform HF solution. The domain-wall calculations presented in the main text are performed within the spin–valley polarized quarter-metal (SVQM) phase, where two time-reversal-related ferromagnetic ground states are exactly degenerate and serve as the boundary conditions for the domain-wall calculations.

\section{A. Hartree--Fock mean-field theory of uniform ground state}

We consider rhombohedral bilayer graphene (rBG) in the presence of an interlayer potential $U_D$ controlled by a perpendicular displacement field, which isolates the low-energy bands near charge neutrality. The electronic structure is described within the Slonczewski--Weiss--McClure (SWMc) continuum model obtained by expanding the tight-binding Hamiltonian around the two graphene valleys, $K$ and $K'$. Throughout this work, we use the same microscopic parameters as in the main text.

The graphene lattice has lattice constant $a=2.46~\mathring{\mathrm{A}}$, nearest-neighbor carbon bond length $a_{\rm CC}=1.42~\mathring{\mathrm{A}}$, and interlayer spacing $d_{\rm layer}=3.35~\mathring{\mathrm{A}}$. Together with the SWMc hopping amplitudes listed in Table~\ref{tab:graphene_params_BBG}, these structural parameters determine the non-interacting band structure and the interaction-driven phase diagram discussed below.

We consider a single $p_z$ orbital on each carbon atom. In the continuum description, the low-energy states are labeled by valley $\tau=\pm1$, spin $s=\uparrow,\downarrow$, layer $l=1,2$, and sublattice $\sigma=A,B$. In the layer--sublattice basis, we define the four-component annihilation spinor
\begin{align}
c_{\tau,s}(\vec k)=
\begin{pmatrix}
c_{\tau,s,A_1}(\vec k)\\
c_{\tau,s,B_1}(\vec k)\\
c_{\tau,s,A_2}(\vec k)\\
c_{\tau,s,B_2}(\vec k)
\end{pmatrix},
\end{align}
where $\vec k$ is measured relative to the center of valley $\tau$.

The mean-field Hamiltonian is
\begin{align}
\hat H^{\rm MF}(\bm{k})
=
\hat H^{\rm SWMc}(\bm{k})
+\hat H^{\rm SOC}
+\hat \Sigma^{\rm HF,C}(\bm{k})
+\hat \Sigma^{\rm HF,Hund},
\label{eq:MF_uniform}
\end{align}
where $\hat H^{\rm SWMc}(\bm{k})$ is the SWMc band Hamiltonian, $\hat H^{\rm SOC}$ is the intrinsic Kane--Mele spin--orbit coupling (SOC), $\hat \Sigma^{\rm HF,C}(\bm{k})$ is the Hartree--Fock self-energy arising from the long-range Coulomb interaction, and $\hat \Sigma^{\rm HF,Hund}$ is the Hartree--Fock self-energy arising from the intervalley Hund's interaction. The individual terms are defined in the following subsections.

\subsection{Band Hamiltonian and Spin-orbit Interaction}
The continuum band Hamiltonian is diagonal in the spin and valley degrees of freedom and can be written as
\begin{align}
\hat{\mathcal H}^{\rm SWMc}
=
\sum_{\vec{k},\tau,s}
c_{\tau s}^{\dagger}(\vec{k})
\,h_\tau(\vec{k})\,
c_{\tau s}(\vec{k}),
\end{align}
where the $4\times4$ matrix $h_\tau(\vec k)$ is the valley-resolved SWMc Hamiltonian acting on the layer--sublattice space. Explicitly, the valley-resolved SWMc Hamiltonian is given by
\begin{align}
    h_\tau(\vec k)=\begin{bmatrix}
        t(\vec k)+U_1 & t_{12}(\vec k)\\
        t^\dagger_{12}(\vec k)& t(\vec k)+U_2
    \end{bmatrix}_{4\times 4}
\end{align}
The Hamiltonian consists of intra-layer hopping ($t$), nearest-layer hopping ($t_{12}$), and layer-dependent onsite potentials $U_i$ arising from externally applied gate voltages and/or broken symmetries. Explicitly,
\begin{align}
    t(\vec{k}) = \begin{bmatrix}
      0 & v_0 \pi^\dagger \\
      v_0 \pi & 0
    \end{bmatrix}, \quad 
    t_{12}(\vec{k}) = \begin{bmatrix}
      -v_4 \pi^\dagger & v_3 \pi \\
      \gamma_1 & -v_4 \pi^\dagger
    \end{bmatrix}
\end{align}
where $\pi = \tau k_x+i k_y$ is a linear momentum, and $\gamma_i$ ($i=0,\dots, 4$) are hopping amplitudes of the tight-binding model with corresponding velocity parameters $v_i=(\sqrt{3}/2)a\gamma_i/\hbar$. The ${U_i}$ terms are
\begin{align}
    U_1 = \begin{bmatrix}
      \frac{\Delta+\delta+U_D}{2} & 0 \\
      0 & \frac{\Delta+U_D}{2}
    \end{bmatrix}, \quad
    U_2 = \begin{bmatrix}
      \frac{\Delta-U_D}{2} & 0 \\
      0 & \frac{\Delta+\delta-U_D}{2}
    \end{bmatrix}
    \label{eq:Ui}
\end{align}

We use the model parameters listed in Table I, which are chosen to match quantum oscillation frequency signatures in Ref.~\cite{zhou2021half}. The tunable parameter $U_D$ in Eq.~\ref{eq:Ui}  is the interlayer potential, and is proportional to out-of-plane electric displacement field. 
\begin{table}[h]
\caption{\label{tab:graphene_params_BBG}Tight-binding parameters (in meV) for rhombohedral graphene, see also Refs.~\cite{zhou2021half}.}
\begin{tabular}{|l|l|l|l|l|l|l|l|}
\hline
$\gamma_0$ & $\gamma_1$ & $\gamma_2$ & $\gamma_3$ & $\gamma_4$  & $\Delta$ & $\delta$ & $U_D$\\
\hline
$3160$ & $380$ & $-15$ & $-290$ & $141$ &  $-23$ & $-10.5$ & $25$ \\
\hline
\end{tabular} 
\end{table}


In the large-displacement-field regime considered here, the low-energy
electronic states are predominantly polarized onto the non-dimer
$A_1$ and $B_2$ sublattices, suppressing the leading effect of Rashba
SOC. The dominant SOC contribution is therefore the
momentum-independent intrinsic Kane--Mele term
\begin{align}
    \hat H^{\rm SOC}
    =-\frac{\lambda_{\rm soc}}{2}
    \hat\sigma_z\hat\tau_z\hat s_z .
\end{align}
Here, $\hat\sigma_z$, $\hat\tau_z$, and $\hat s_z$ are Pauli
matrices acting in sublattice, valley, and spin space, respectively.
We take $\sigma_z=+1$ ($-1$) on the $A$ ($B$) sublattice,
$\tau_z=+1$ ($-1$) in the $K$ ($K'$) valley, and
$s_z=+1$ ($-1$) for spin $\uparrow$ ($\downarrow$).
For the sign of $U_D$ and the hole-doped regime considered here, the
relevant valence-band states are predominantly polarized onto the
$B_2$ sublattice. With the convention $\lambda_{\rm soc}>0$, the SOC
term above therefore favors the $K\uparrow$ and $K'\downarrow$ hole
flavors. We use $\lambda_{\rm soc}=0.1~{\rm meV}$, consistent with
the experimentally estimated intrinsic SOC strength in rhombohedral
graphene~\cite{auerbach2025isospin}. The opposite sign convention would simply interchange the two
SOC-favored spin--valley pairs,
$\{K\uparrow,K'\downarrow\}\leftrightarrow
\{K\downarrow,K'\uparrow\}$.
In the absence of an external Zeeman field, this relabeling does not
affect the existence, structure, or transition between the two- and
four-component domain walls studied here.

\subsection{Electron--electron interactions}

The electron--electron interaction consists of two contributions: a long-range screened Coulomb interaction and a lattice-scale intervalley Hund's interaction. The former drives exchange ferromagnetism, while the latter favors parallel spin alignment between opposite valleys and is responsible for the transition between the two- and four-component intervalley-coherent states discussed below. Both interactions are treated within the Hartree--Fock approximation.

The long-range Coulomb interaction is described by
\begin{align}\label{eq:H_Coulomb}
\hat{\mathcal H}^{\rm C}
=
\frac{1}{2A}
\sum_{\vec q}
V_C(\vec q)
:\hat n(\vec q)\hat n(-\vec q):~,\quad 
\hat{n}(\vec q) = \sum_{\vec{k},\tau,s,l,\sigma} c_{\tau,s,l,\sigma}^\dagger(\vec{k}+\vec{q})  c_{\tau,s,l,\sigma}(\vec{k}).
\end{align}
Here $::$ denotes normal ordering and $A$ is the sample area. The screened Coulomb potential is taken to be
$V_C(\vec q)=2\pi k_e\tanh(|\vec q|d)/(\epsilon_r|\vec q|)$,
where $k_e=1.44~\mathrm{eV\,nm}$ is the Coulomb constant, $\epsilon_r=15$ is the effective dielectric constant, and $d=50~\mathrm{nm}$ is the gate distance.

The lattice-scale intervalley Hund's interaction originates from short-range exchange processes between electrons in opposite valleys and is modeled by
\begin{align}\label{eq:H_Hund}
\hat{\mathcal H}^{\rm Hund}
=\frac{g_\perp}{2A} \sum_{\tau}\sum_{\vec k, \vec k'} \sum_{\vec q}  \sum_{l,\sigma} \sum_{s ,s'}
: c^\dagger_{\bar\tau s l\sigma}(\vec k)\, c_{\tau sl\sigma}(\vec k+ \vec q)\,
  c^\dagger_{\tau s'l\sigma}(\vec k')\, c_{\bar\tau s'l\sigma}(\vec k' - \vec q) :,
\end{align}
where $g_\perp>0$ is the intervalley Hund's coupling strength, and $\bar{\tau}=-\tau$ denotes the opposite valley. A positive $g_\perp$ favors ferromagnetic alignment of spins between opposite valleys.

The above intervalley Hund's coupling $\hat{\mathcal H}^{\rm Hund}$ is one of the many two-body interactions allowed by the lattice symmetries of graphene. Because these interactions originate from processes involving momenta of order the inverse lattice spacing, they are short-ranged on the scale of the continuum theory and can, to leading order be represented by contact interactions. Corrections beyond the zero-range approximation were studied in Ref.~\cite{de2023global}. The symmetry classification of these interactions was developed in Ref.~\cite{aleiner2007spontaneous} for monolayer graphene and in Ref.~\cite{lemonik2010spontaneous} for bilayer graphene. Their bare coupling constants were calculated in Ref.~\cite{wei2025landau} by approximating the three-dimensional electronic wavefunctions as linear combinations of carbon $2p_z$ orbitals with an effective nuclear charge, and their RG  flow was studied in Ref.~\cite{kharitonov2012phase,PhysRevB.111.125127}.

The above analyses were originally developed for monolayer graphene and bilayer graphene. Closely related intervalley Hund interactions have also been incorporated into continuum models of rhombohedral multilayer graphene~\cite{PhysRevB.110.245118,chatterjee2022inter}. The intrasublattice component presents no difficulty and can be represented by a momentum-independent local interaction. For the intersublattice component, however, Ref.~\cite{PhysRevB.110.245118} constructs a $C_3$-invariant interaction using the phenomenological kernel $V(\bm q)
=J_H e^{2i \tau \theta_{\bm q}}$,
where $\theta_{\bm q}=\arg(q_x+i q_y)$. Since it has no direction-independent limit when $\bm q\rightarrow0$, the potential is nonanalytic at $\bm q=0$. As a result, its Fourier transform  has a slow algebraic decay in real space $e^{im\theta_{r}}/r^2$, and therefore does not correspond to an ultrashort-range potential in the continuum model.
For simplicity, we therefore adopt a minimal local description in which the intervalley Hund interaction is diagonal in the microscopic sublattice index and has the same coupling strength on every graphene layer. This approximation is physically motivated when the carbon $2p_z$ orbitals are sufficiently compact compared with the interlayer separation. The leading layer-dependent interaction is the intralayer and interlayer Coulomb energies difference, see Ref.~~\cite{kolavr2026electrostatically} and Section.~VI of \cite{wolf2024magnetism}.


Within the Hartree--Fock approximation, the quartic interaction terms are factorized with respect to the single-particle density matrix, giving the Coulomb and intervalley Hund self-energies entering Eq.~(\ref{eq:MF_uniform}). The resulting Coulomb and intervalley Hund Hartree--Fock self-energies entering Eq.~\ref{eq:MF_uniform} are given below. Applying the Hartree--Fock decoupling to Eq.~\ref{eq:H_Coulomb} gives
\[
\hat \Sigma^{\rm HF,C}(\bm{k})
=
\hat \Sigma^{\rm H,C}
+
\hat \Sigma^{\rm F,C}(\bm{k}),
\]
\begin{align}
    \Sigma^{\rm H,C}_{\alpha,\beta} =\frac{N}{A}\delta_{\alpha,\beta} V_C(0)\quad,\quad
    \Sigma^{\rm F,C}_{\alpha,\beta}(\bm{k}) =-\frac{1}{A}\sum_{q}V_C(\vec q)\rho_{\alpha,\beta}(\vec k+\vec q).
\end{align}
Here $N$ is total number of occupied electronic states, and $\rho_{\alpha,\beta}(\vec k)=\braket{c^\dagger_\beta(\vec k) c_\alpha(\vec k)}$ is the single-particle density matrix. The composite indices $\alpha,\beta$ label the combined valley $(\tau)$, spin $(s)$, layer $(l)$, and sublattice $(\sigma)$ degrees of freedom. The Hartree term describes the electrostatic interaction generated by the total charge density, whereas the Fock term gives the exchange interaction responsible for spontaneous symmetry breaking. For the dual-gated Coulomb interaction, the $\vec q=0$ Hartree contribution is finite and depends only on the spatially averaged carrier density. At fixed carrier density, it produces a uniform shift of the mean-field potential and contributes equally to the competing states. We therefore omit the $\vec q=0$ Hartree term from the self-consistent mean-field Hamiltonian, while retaining all finite-$\vec q$ Hartree components associated with spatial charge redistribution.

The corresponding intervalley Hund Hartree--Fock self-energy is
\[
\hat\Sigma^{\rm HF,Hund}
=
\hat\Sigma^{\rm H,Hund}
+
\hat\Sigma^{\rm F,Hund},
\]
\begin{align}
    \hat\Sigma^{\rm H,Hund}&=\frac{g_\perp}{A} \sum_{\tau,l,\sigma}
\sum_{\vec k,s}
c^\dagger_{\bar\tau sl\sigma}(\vec k)\,
c_{\tau sl\sigma}(\vec k)\sum_{\vec k',s'}
\left\langle
c^\dagger_{\tau s'l\sigma}(\vec k')\,
c_{\bar\tau s'l\sigma}(\vec k')
\right\rangle \qquad \text{(Hartree: $\vec q=0$)}\label{eq:Hund_H_uniform}\\
\hat\Sigma^{\rm F,Hund}&=-\frac{g_\perp}{A}\sum_{\tau,l,\sigma} \sum_{\vec k,s ,s'} c^\dagger_{\bar\tau s l\sigma}(\vec k)\, c_{\bar\tau s'l\sigma}(\vec k)\sum_{\vec k'}\braket{c^\dagger_{\tau s'l\sigma}(\vec k')\, c_{\tau sl\sigma}(\vec k')}\qquad \text{(Fock: $\vec q=\vec k'-\vec k$)} \label{eq:Hund_F_uniform}
\end{align}
The Hartree contribution penalizes spin-preserving intervalley coherence, whereas the Fock contribution lowers the energy through intervalley exchange and favors parallel spin alignment between opposite valleys. The competition between these two terms determines whether the intervalley-coherent state remains confined to two spin-valley flavors or reconstructs into a four-component state.

The self-consistent Hartree--Fock equations are solved over the $(n_e,U_D)$ phase space to determine the uniform quarter-metal ground states. We then examine how increasing intervalley Hund's coupling drives a continuous transition from the two-component to the four-component intervalley-coherent phase, which provides the microscopic origin of the analogous domain-wall transition discussed in the main text.

\subsection{Uniform Quarter-Metal Ground States}
\begin{figure}
    \centering
    \includegraphics[width=\linewidth]{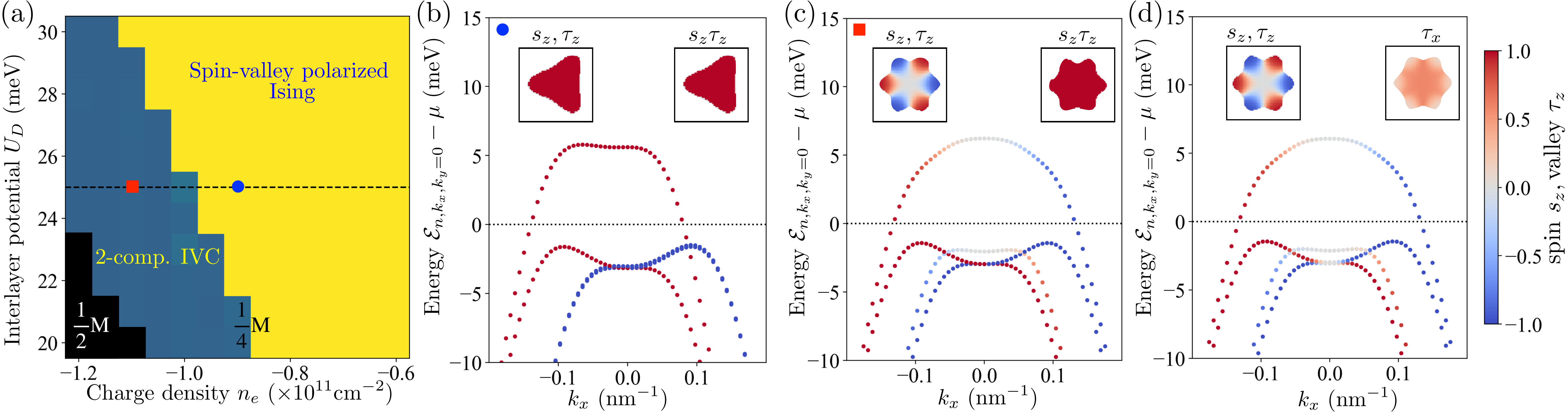}
    \caption{(a) Hartree--Fock phase diagram as a function of carrier density $n_e$ and interlayer potential $U_D$ at $g_\perp=0$. The horizontal dashed line marks the fixed interlayer potential $U_D=25$ meV used throughout this work. The system realizes spin--valley polarized and intervalley-coherent (IVC) quarter-metal phases, while the black region denotes the half-metal ($1/2$M) phase. The blue dot marks the parameter set used for the domain-wall calculations in the main text, and the red square indicates the density used to study the Hund-driven transition within the IVC quarter metal.
(b) Quasiparticle band structure of the spin--valley quarter metal (SVQM) at $n_e=-0.09\times10^{12}\,\mathrm{cm}^{-2}$ along $k_y=0$. The quasiparticle energies are measured relative to the chemical potential $\mu$, and the horizontal dashed line marks the Fermi level.
(c) Quasiparticle band structure of the two-component IVC quarter metal at $n_e=-0.11\times10^{12}\,\mathrm{cm}^{-2}$ and $g_\perp=0$. The insets show the spin and valley order parameters evaluated over the occupied Fermi sea. The IVC order is confined to the $K\uparrow$ and $K'\downarrow$ flavors, giving $s_z\tau_z=1$ throughout the occupied Fermi sea.
(d) Quasiparticle band structure at the same carrier density for $g_\perp|n_e|/\lambda_{\rm soc}=5.5$, corresponding to the four-component IVC quarter metal. The left inset shows the spin and valley polarizations, while the right inset shows the intervalley coherence $\tau_x$ evaluated over the occupied Fermi sea.}
    \label{fig:QM_phases}
\end{figure}

We now solve the self-consistent Hartree--Fock equations for the uniform system as a function of carrier density $n_e$ and interlayer potential $U_D$. Figure~\ref{fig:QM_phases}(a) shows the resulting phase diagram at $g_\perp=0$. Depending on the carrier density and displacement field, the system realizes several symmetry-broken metallic phases distinguished by their spin and valley polarization. The horizontal dashed line denotes the fixed interlayer potential $U_D=25$ meV used throughout this work, along which the representative band structures in Figs.~\ref{fig:QM_phases}(b)--(d) are obtained. Throughout the main text, we focus on the spin--valley polarized quarter-metal (SVQM) phase at $n_e=-0.09\times10^{12}\ {\rm cm}^{-2}$, indicated by the blue dot in Fig.~\ref{fig:QM_phases}(a), where two time-reversal-related ferromagnetic ground states are exactly degenerate. These degenerate states provide the boundary conditions for the domain-wall calculations discussed in the main text.

Representative quasiparticle band structures of the metallic phases are shown in Fig.~\ref{fig:QM_phases}(b)--(d). Figure~\ref{fig:QM_phases}(b) shows the SVQM band structure at this carrier density. The band colors indicate the spin and valley polarizations, $s_z$ and $\tau_z$, and the horizontal dashed line indicates the Fermi level. Figures~\ref{fig:QM_phases}(c) and (d) instead consider a slightly higher hole density at $n_e=-0.11\times10^{12}\ {\rm cm}^{-2}$ indicated by the red square in Fig.~\ref{fig:QM_phases}(a),
where the ground state is an intervalley-coherent (IVC) quarter metal. In the absence of intervalley Hund's coupling, the IVC order resides within two spin--valley flavors, $K\uparrow$ and $K'\downarrow$, giving rise to the two-component IVC state shown in Fig.~\ref{fig:QM_phases}(c). The inset shows the momentum distribution of order parameters inside the Fermi surface. For 2-component IVC phase, $s_z\tau_z$ order parameter is $1$ throughout the Fermi surface, while $s_z$ and $\tau_z$ oscillates around the Fermi surface. Upon increasing the Hund's coupling to $g_\perp|n_e|/\lambda_{\rm soc}=5.5$, the quasiparticle spectrum evolves into the four-component IVC state shown in Fig.~\ref{fig:QM_phases}(d).

To make the distinction between the two IVC states explicit, an
occupied HF quarter-metal state can be written as
\begin{align}
\ket{\psi_{\vec k}}
&=
\sum_{\tau,s}
c^\dagger_{\tau s}(\vec k)u_{\tau s}(\vec k)\,
\ket{\mathrm{CN}},
\label{eq:uniform_spin_valley_spinor}
\end{align}
where $\ket{\mathrm{CN}}$ denotes the charge-neutral reference state
and each $u_{\tau s}(\vec k)$ is a four-component layer--sublattice
spinor. Here, $c^\dagger_{\tau s}(\vec k)$ creates the relevant
electron or hole carrier relative to $\ket{\mathrm{CN}}$. The product
$c^\dagger_{\tau s}u_{\tau s}
=\sum_{l,\sigma}c^\dagger_{\tau sl\sigma}z_{\tau sl\sigma}$
implicitly contracts the layer--sublattice indices. The
many-body HF ground state is obtained by occupying all such states
within the Fermi sea.

The transverse spin polarization and spin-preserving intervalley
coherence are
\begin{align}
s_x
&=
\frac{1}{N_e}
\sum_{\vec k,\tau}
\left[
\braket{
c^\dagger_{\tau\uparrow}(\vec k)
c_{\tau\downarrow}(\vec k)}
+\mathrm{c.c.}
\right],
\label{eq:uniform_sx}
\\
\tau_x
&=
\frac{1}{N_e}
\sum_{\vec k,s}
\left[
\braket{
c^\dagger_{Ks}(\vec k)
c_{K's}(\vec k)}
+\mathrm{c.c.}
\right],
\label{eq:uniform_taux}
\end{align}
where $N_e=|n_e|A$, and the layer--sublattice contraction is implicit.
In terms of the spinor components,
\begin{align}
s_x(\vec k)
&\propto
2\operatorname{Re}\!\left[
u_{K\uparrow}^*(\vec k)u_{K\downarrow}(\vec k)
+
u_{K'\uparrow}^*(\vec k)u_{K'\downarrow}(\vec k)
\right],
\\
\tau_x(\vec k)
&\propto
2\operatorname{Re}\!\left[
u_{K\uparrow}^*(\vec k)u_{K'\uparrow}(\vec k)
+
u_{K\downarrow}^*(\vec k)u_{K'\downarrow}(\vec k)
\right].
\end{align}

For the two-component IVC state,
$u_{K'\uparrow}=u_{K\downarrow}=0$, so that
$s_x=\tau_x=0$. Its coherence between $K\uparrow$ and
$K'\downarrow$ instead lies in a mixed spin--valley channel. In the
self-consistent four-component state, both $u_{K'\uparrow}$ and
$u_{K\downarrow}$ become nonzero, allowing finite $s_x$ and
$\tau_x$, as seen in the insets of Fig.~\ref{fig:QM_phases}(d).
Thus, these transverse order parameters diagnose the reconstruction
out of the original $\{K\uparrow,K'\downarrow\}$ subspace. Throughout
this evolution, the system remains a metallic quarter metal, while
its internal spin--valley structure changes continuously.

\begin{figure}[t]
    \centering
    \includegraphics[width=0.35\linewidth]{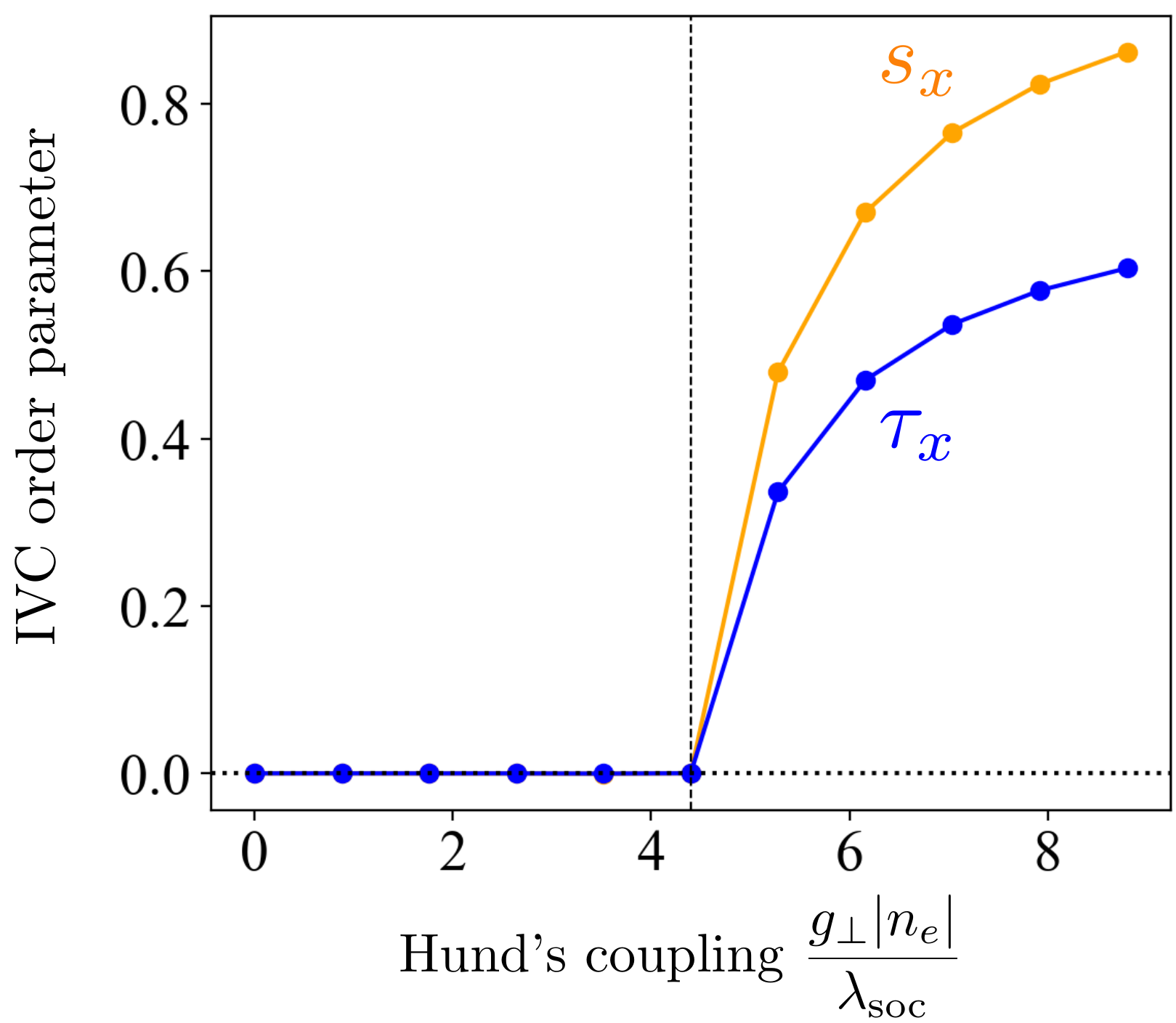}
    \caption{Evolution of the spin and valley order parameters as a function of the dimensionless intervalley Hund's coupling $g_\perp|n_e|/\lambda_{\rm soc}$ at $n_e=-0.11\times10^{12}\,\mathrm{cm}^{-2}$. The in-plane spin and valley components develop continuously above the critical coupling $g_\perp^{*}|n_e|/\lambda_{\rm soc}=4.5$, signaling a second-order transition from the two-component to the four-component IVC quarter-metal phase.}
    \label{fig:phase_transition}
\end{figure}

To characterize the continuous onset of these additional spin--valley
components, Fig.~\ref{fig:phase_transition} shows the evolution of the spin and valley order parameters as a function of the dimensionless Hund's coupling $g_\perp|n_e|/\lambda_{\rm soc}$ at $n_e=-0.11\times10^{12}\ {\rm cm}^{-2}$. Below the critical coupling, the in-plane spin polarization and intervalley coherence vanish, corresponding to the two-component IVC state. At the critical coupling,
$g_\perp^*|n_e|/\lambda_{\rm soc}=4.5$,
both order parameters develop continuously from zero, signaling a second-order transition into the four-component IVC phase. The continuous onset of these additional order parameters demonstrates that the competition between the long-range Coulomb interaction and the intervalley Hund's coupling naturally drives a reconstruction of the IVC order.

This result provides the microscopic context for the domain-wall calculations presented in the main text. Since the center of the domain wall locally realizes an intervalley-coherent state, the continuous transition between the two- and four-component IVC phases in the uniform system provides a natural basis for understanding the analogous reconstruction between the two- and four-component domain walls induced by increasing intervalley Hund's coupling.

\section{B. Hartree--Fock Mean--Field Theory of Spin--Valley Domain Walls}

In this section, we extend the uniform Hartree--Fock theory to describe spin--valley domain walls. We first formulate the one-dimensional magnetic-supercell geometry, then derive the Hartree--Fock self-consistent equations in the folded momentum basis, and finally discuss the numerical convergence and microscopic properties of the resulting domain-wall solutions.

\subsection{One-dimensional Magnetic Supercell}

We consider domain walls separating two time-reversal-related spin--valley polarized quarter-metal (SVQM) ground states. The calculation is performed in a single periodic one-dimensional magnetic supercell of length $L$. Translational symmetry is preserved along the $y$ direction, while the spin--valley order parameter varies periodically along $x$. Consequently, each supercell contains two equivalent domain walls. The in-plane
coordinate axes are chosen such that $\vec e_x$ and $\vec e_y$ are the
unit vectors parallel and perpendicular, respectively, to the graphene
Dirac-point momentum $\vec K$ measured from the $\Gamma$ point.

The broken translational symmetry along $x$ couples momentum states differing by reciprocal lattice vectors of the magnetic supercell,
$ 
G=\frac{2\pi m}{L},$
where $m\in\mathbb{Z}$. The momentum along the domain wall, $k_y$, remains a good quantum number, whereas the momentum dependence along $x$ is represented by the discrete reciprocal-vector indices $G$. The single-particle basis is therefore labeled by $(k_y,G)$, with the physical momentum entering the continuum band Hamiltonian given by $\vec{k}_G=(G,k_y)$.

\subsection{Mean-Field Hamiltonian in the Magnetic Supercell}

In the one-dimensional magnetic-supercell geometry, the single-particle basis states are labeled by $(G,\alpha)$, where $\alpha=\{s,\tau,l,\sigma\}$ collectively denotes the spin, valley, layer, and sublattice degrees of freedom. The momentum $k_y$ along the domain wall remains conserved, while the momentum along $x$ is represented by the reciprocal supercell index $G$. In this basis, the mean-field Hamiltonian takes the form given in Eq.~\ref{eq:HMF_DW} of the main text,
\begin{align}
\hat H^{\rm MF}_{G\alpha,G'\beta}(k_y)
=
\delta_{GG'}
\left[
\hat H^{\rm SWMc}(G\vec e_x+k_y\vec e_y)
+\hat H^{\rm SOC}
\right]_{\alpha\beta}
+
\hat\Sigma^{\rm HF,C}_{G\alpha,G'\beta}(k_y)
+
\hat\Sigma^{\rm HF,Hund}_{G\alpha,G'\beta}(k_y).
\label{eq:HMF_DW_copy}
\end{align}
The off-diagonal matrix elements with $G\neq G'$ arise from the spatially varying charge and spin--valley order parameters and describe the hybridization between different reciprocal components of the magnetic supercell.

The Coulomb and intervalley Hund self-energies are obtained using the same Hartree--Fock decoupling introduced in Sec.~A. In the magnetic supercell, however, the single-particle density matrix also carries reciprocal-vector indices,
\begin{align}
\rho_{G\alpha,G'\beta}(k_y)
=
\left\langle
c^\dagger_{G'\beta}(k_y)
c_{G\alpha}(k_y)
\right\rangle.
\label{eq:rho_DW}
\end{align}
Because translational symmetry is preserved along $y$, the density matrix is diagonal in $k_y$ but may be off diagonal in $G$.
The Coulomb Hartree and Fock self-energies are
\begin{align}
\Sigma^{\rm H,C}_{G\alpha,G'\beta}
&=
\delta_{\alpha\beta}
\frac{1}{A}
\sum_{k_y',G'',\gamma}
V_C\!\left[(G-G')\vec e_x\right]
\rho_{G''\gamma,G''+G-G'\gamma}(k_y'),
\label{eq:Hartree_DW}
\\
\Sigma^{\rm F,C}_{G\alpha,G'\beta}(k_y)
&=
-\frac{1}{A}
\sum_{k_y'}\sum_{G_r}
V_C\!\left[
G_r\vec e_x+(k_y-k_y')\vec e_y
\right]
\rho_{G+G_r,\alpha;\,G'+G_r,\beta}(k_y').
\label{eq:Fock_DW}
\end{align}
Here $G_r=2\pi m_r/L$ is the reciprocal momentum transferred along the $x$ direction. The Hartree self-energy is independent of $k_y$ because the charge density is uniform along $y$. The spatially uniform component, corresponding to $G=G'$, is omitted from the mean-field Hamiltonian. In contrast, the finite-$G-G'$ Hartree components describe the electrostatic energy associated with charge redistribution across the domain wall.

Applying the same decoupling to the intervalley Hund interaction gives
\begin{align}
\hat\Sigma^{\rm H,Hund}
={}&
\frac{g_\perp}{A}
\sum_{\tau,l,\sigma}
\sum_{s,s'}
\sum_{G_r}
\sum_{k_y,G}
c^\dagger_{\bar\tau s l\sigma,G}(k_y)\,
c_{\tau s l\sigma,G+G_r}(k_y)
\nonumber\\
&\hspace{1.0cm}\times
\sum_{k_y',G'}
\left\langle
c^\dagger_{\tau s'l\sigma,G'}(k_y')\,
c_{\bar\tau s'l\sigma,G'-G_r}(k_y')
\right\rangle ,
\label{eq:Hund_H_DW}
\\[0.3cm]
\hat\Sigma^{\rm F,Hund}
={}&
-\frac{g_\perp}{A}
\sum_{\tau,l,\sigma}
\sum_{s,s'}
\sum_{G_r}
\sum_{k_y,G}
c^\dagger_{\bar\tau s l\sigma,G}(k_y)\,
c_{\bar\tau s'l\sigma,G+G_r}(k_y)
\nonumber\\
&\hspace{1.0cm}\times
\sum_{k_y',G'}
\left\langle
c^\dagger_{\tau s'l\sigma,G'}(k_y')\,
c_{\tau s l\sigma,G'-G_r}(k_y')
\right\rangle .
\label{eq:Hund_F_DW}
\end{align}
The Hund Hartree contribution couples spin-preserving intervalley-coherent components, whereas the Hund Fock contribution couples the valley-resolved spin density matrices and favors ferromagnetic spin alignment between opposite valleys. The reciprocal-lattice transfer $G_r$ allows both contributions to vary spatially across the magnetic supercell.

Starting from an initial spin--valley texture satisfying the prescribed domain-wall boundary conditions, we diagonalize Eq.~\ref{eq:HMF_DW_copy}, reconstruct the density matrix in Eq.~\ref{eq:rho_DW}, and update the Hartree--Fock self-energies iteratively until self-consistency is reached. The resulting metastable solutions determine the domain-wall profiles and their excess line energies. In the following subsection, we examine the convergence of these quantities with the magnetic-supercell length.

\subsection{Convergence with Magnetic-Supercell Length}

The converged self-consistent solution determines the domain-wall energetics through the Hartree--Fock mean-field energy density,
\begin{align}
E_{\rm DW}
=
\frac{1}{2A}
\sum_{k_y}
{\rm Tr}
\left[
\hat\rho(k_y)
\left(
\hat H^{\rm MF}(k_y)
+
\hat H^{\rm SWMc}(k_y)+\hat H^{\rm SOC}
\right)
\right],
\label{eq:DW_energy_sup}
\end{align}
where the trace is taken over the reciprocal-vector and internal spin--valley basis indices. The corresponding uniform ground-state energy density is
\begin{align}
E_0
=
\frac{1}{2A}
\sum_{\vec k}^{\rm BZ}
{\rm Tr}
\left[
\hat\rho_0(\vec k)
\left(
\hat H^{\rm MF}_0(\vec k)
+
\hat H^{\rm SWMc}(\vec k)+\hat H^{\rm SOC}
\right)
\right],
\label{eq:uniform_energy_sup}
\end{align}
where $\hat\rho_0$ and $\hat H^{\rm MF}_0$ denote the self-consistent density matrix and mean-field Hamiltonian of the corresponding uniform SVQM ground state. The excess domain-wall line energy is then defined as
\begin{align}
\sigma_{\rm DW}
=
\frac{L}{2}
\left(
E_{\rm DW}
-
E_0
\right),
\label{eq:line_energy_sup}
\end{align}
where the factor of two accounts for the two equivalent domain walls contained within each magnetic supercell.
\begin{figure}[t]
    \centering
    \includegraphics[width=0.35\linewidth]{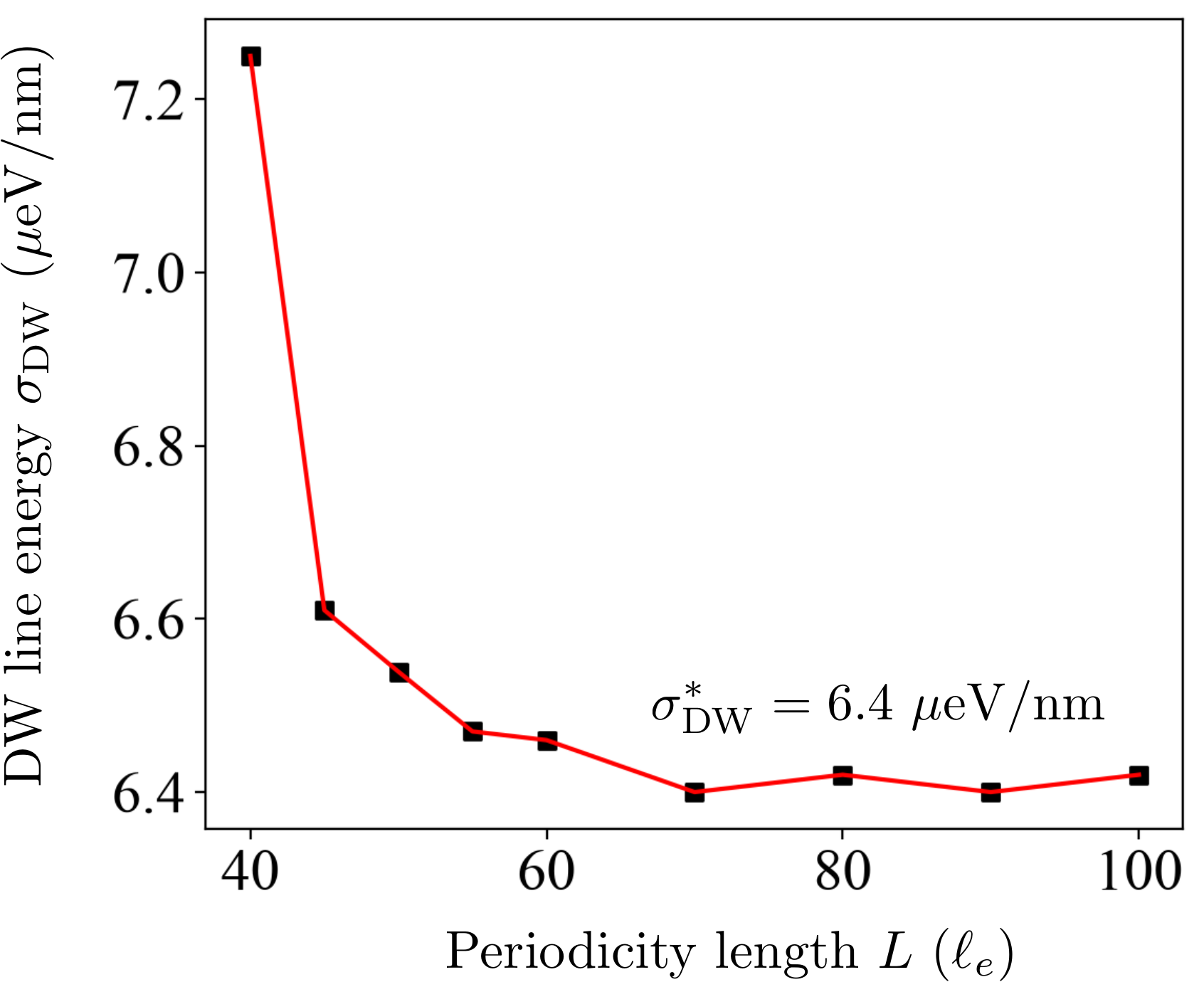}
    \caption{Domain-wall line energy $\sigma_{\rm DW}$ as a function of the magnetic-supercell length $L$. The line energy converges to $\sigma_{\rm DW}^*\simeq6.4~\mu{\rm eV}/{\rm nm}$ for $L\gtrsim70\,\ell_e$, indicating negligible interaction between the two domain walls contained within the supercell. Throughout this work we use $L=160\,\ell_e$.}
    \label{fig:DW_energetics}
\end{figure}
The domain-wall calculations are performed on a discretized one-dimensional Brillouin zone containing $N_{k_y}=21$ momentum points along the translationally invariant $y$ direction. The reciprocal-space expansion retains $N_G$ reciprocal vectors, corresponding to a momentum cutoff $G_{\rm max}=3.4\,\ell_e^{-1}$. The resulting momentum discretization corresponds to an effective supercell area of $A=10^4\,\ell_e^2$.
Figure~\ref{fig:DW_energetics} shows the convergence of the domain-wall line energy with increasing magnetic-supercell length. As the separation between the two domain walls increases, their mutual interaction becomes negligible and the line energy rapidly approaches a constant value. Throughout this work we use $L=160\,\ell_e$, which lies well within the converged regime.
\subsection{Quasiparticle Wavefunctions in the Domain Wall}

\begin{figure}[t]
    \centering
    \includegraphics[width=0.8\linewidth]{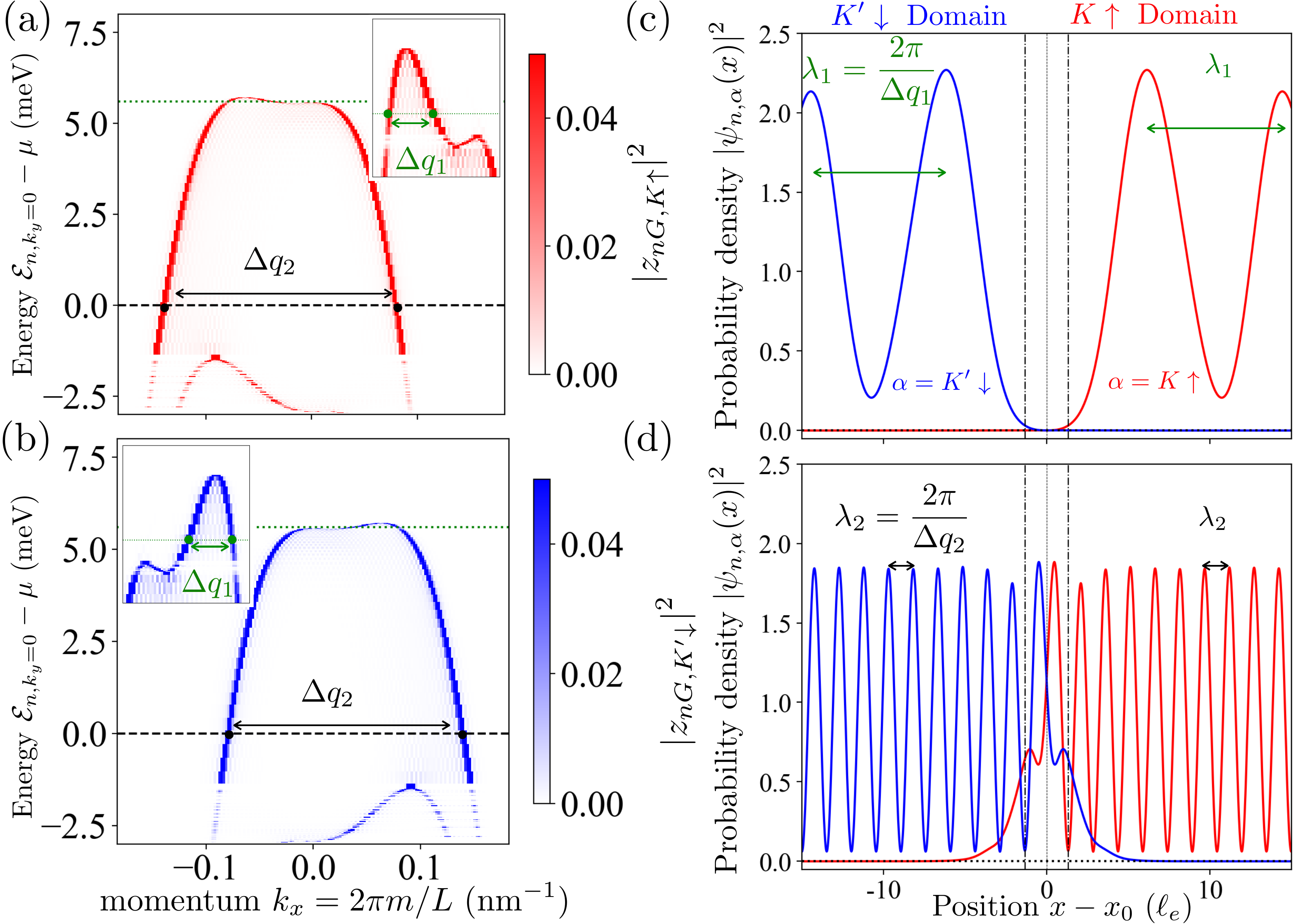}
    \caption{(a),(b) Reciprocal-space probability distributions $|z_{n,K\uparrow}(k_x,k_y=0)|^2$ and $|z_{n,K'\downarrow}(k_x,k_y=0)|^2$, respectively, as functions of the discrete momentum $k_x=2\pi m/L$ and quasiparticle energy $\mathcal{E}_n(k_y=0)-\mu$. The color scale represents the probability weight of each reciprocal component. Insets magnify the energy window $5.5\le\mathcal{E}_n-\mu\le6.0$ meV. The horizontal green and black dashed lines indicate the quasiparticle states shown in (c) and (d), respectively. (c),(d) Real-space probability distributions $|\psi_{n,\alpha}(x)|^2$ of the corresponding quasiparticle states at $\mathcal{E}_n(k_y=0)-\mu=5.6$ meV and at the Fermi level. The vertical dot-dashed lines denote the domain-wall boundaries. The bulk probability density exhibits interference oscillations with wavelength $\lambda=2\pi/\Delta q$, where $\Delta q$ is the momentum separation between the dominant reciprocal components. The oscillation periods satisfy $N=9$ in (c) and $N=51$ in (d). While the finite-energy state is suppressed at the domain-wall boundary, the Fermi-level state penetrates into the domain-wall region.}
    \label{fig:psi_x_supp}
\end{figure}

To gain microscopic insight into the domain-wall electronic structure, we examine the quasiparticle wavefunctions obtained from the self-consistent Hartree--Fock Hamiltonian. For a fixed momentum $k_y$, the $n$th quasiparticle eigenstate is expanded in the magnetic-supercell basis as
\begin{align}
|\Psi_n(k_y)\rangle
=
\sum_{k_x,\alpha}
z_{n,k_x,\alpha}(k_y)
|k_x,\alpha;k_y\rangle,
\end{align}
where $k_x=2\pi m/L$ labels the reciprocal momentum of the magnetic supercell and $\alpha=\{s,\tau,l,\sigma\}$ denotes the spin, valley, layer, and sublattice degrees of freedom. We resolve the quasiparticle amplitudes by flavor,
\begin{align}
|z_{n,\tau s}(k_x,k_y)|^2
=
\sum_{l,\sigma}
|z_{n,k_x,\tau s l\sigma}(k_y)|^2,
\end{align}
which is the quantity shown in Figs.~\ref{fig:psi_x_supp}(a) and (b). For this discussion, we consider the 2-component domain wall at $g_\perp=0$.

Figures~\ref{fig:psi_x_supp}(a) and (b) display the reciprocal-space probability distributions $|z_{n,K\uparrow}(k_x,k_y=0)|^2$ and $|z_{n,K'\downarrow}(k_x,k_y=0)|^2$ as functions of $k_x$ and quasiparticle energy $\mathcal{E}_n(k_y=0)-\mu$. The color scale represents the probability weight of each reciprocal component, while the insets magnify the energy window $5.5\text{--}6.0$ meV above the Fermi level. The horizontal green and black dashed lines indicate the two representative quasiparticle states whose real-space probability distributions are shown in Figs.~\ref{fig:psi_x_supp}(c) and (d), respectively.

The corresponding real-space probability density is obtained by Fourier transforming the reciprocal components,
\begin{align}
|\psi_{n,\alpha}(x)|^2
=
\left|
\sum_{k_x}
z_{n,k_x,\alpha}\,
e^{ik_xx}
\right|^2.
\end{align}
When two dominant reciprocal components with momenta
$k_{x,1}$ and $k_{x,2}$ interfere, the probability density
oscillates with wavevector
\begin{align}
\Delta q
=
|k_{x,1}-k_{x,2}|,
\end{align}
corresponding to a real-space wavelength
\begin{align}
\lambda
=
\frac{2\pi}{\Delta q}.
\end{align}
Because the magnetic supercell contains two equivalent bulk regions, each extending over a distance $L/2$ between neighboring domain walls, the oscillation wavelength satisfies
\begin{align}
N\lambda=\frac{L}{2},
\end{align}
where $N$ is the number of oscillation periods between neighboring domain walls. For the representative quasiparticle states shown in
Figs.~\ref{fig:psi_x_supp}(c) and (d), we obtain
$N=9$ at
$\mathcal{E}_n(k_y=0)-\mu=5.6$ meV and
$N=51$ at the Fermi level, corresponding to a substantially larger momentum separation $\Delta q$ between the dominant reciprocal components of the latter state.

Figures~\ref{fig:psi_x_supp}(c) and (d) compare the real-space probability distributions of two representative quasiparticle states. For the state at $\mathcal{E}_n(k_y=0)-\mu=5.6$ meV, the probability density oscillates within the corresponding bulk region while remaining strongly suppressed at the domain-wall boundary. In contrast, the quasiparticle state at the Fermi level exhibits appreciable penetration of the $K\uparrow$ and $K'\downarrow$ components into the domain-wall region. This enhanced penetration demonstrates that the domain wall strongly reconstructs the low-energy quasiparticle wavefunctions, whereas higher-energy bulk states remain largely excluded from the domain-wall center.

\subsection{Domain-Wall Width Estimation}

To quantify the spatial extent of the domain wall, we fit the self-consistent order-parameter profiles to the conventional one-dimensional domain-wall form
\begin{align}
O(x)
=
O_0
\tanh\!\left(\frac{x-x_0}{\xi}\right),
\end{align}
where $O(x)$ denotes the corresponding spin or valley polarization, $x_0$ is the domain-wall center, and $\xi$ is the characteristic width parameter. We define the domain-wall width as
\begin{align}
w=2\xi.
\end{align}

For the two-component domain wall, the spin and valley polarizations remain locked throughout the wall, so fitting either $s_z(x)$ or $\tau_z(x)$ yields the same width. In contrast, the four-component domain wall exhibits a partial relaxation of spin--valley locking within the wall due to the enhanced intervalley Hund's interaction. Consequently, the spin and valley order parameters acquire different characteristic widths, denoted by $w_s$ and $w_\tau$, respectively. We consistently find $w_s>w_\tau$, indicating that the valley polarization varies over a shorter length scale than the spin polarization. This difference reflects the stronger uniaxial anisotropy in the valley sector.

\section{CP$^3$ Long-Wavelength Theory}

To complement the microscopic Hartree--Fock calculations, we develop a continuum CP$^3$ field theory describing the long-wavelength spin--valley textures of the quarter-metal phase. Similar effective theories have been successfully applied to domain walls in magic-angle twisted bilayer graphene~\cite{kwan2021domain} and quantum Hall ferromagnets~\cite{lian20164}. The continuum theory provides a transparent description of the domain-wall reconstruction and enables direct comparison with the self-consistent Hartree--Fock results. The exchange stiffness and anisotropy coefficients entering the free-energy functional are determined directly from the microscopic Hartree--Fock calculations, while the intervalley Hund's coupling is treated as an independent parameter.

A quarter metal is locally described by a normalized four-component spinor $|\psi(x)\rangle$, defined only up to an overall phase. Equivalently, its order parameter is the matrix-valued rank-one projector
\begin{align}
    \hat P(x)=|\psi(x)\rangle\langle\psi(x)|,
    \qquad
    \hat P^2=\hat P,\quad
    \mathrm{Tr}\,\hat P=1,
\end{align}
whose order-parameter manifold is the complex projective space CP$^3$.

\subsection{Free-Energy Functional}

The domain-wall texture is assumed to be translationally invariant along the wall direction ($y$), so that the continuum theory reduces to a one-dimensional problem. The corresponding excess free energy per unit length of the
domain wall, measured relative to the uniform ground state, is
\begin{align}
\mathcal{F}[\hat P]
=
\int dx\,
\mathcal{H}[\hat P],
\label{eq:CP3_energy}
\end{align}
where the local free-energy density is measured relative to
the uniform ground state, so that
$\mathcal H[\hat P_{\rm gs}]=0$:
\begin{align}
\mathcal{H}[\hat P]
=|n_e|\left(
\frac{\rho_s}{2}
\mathrm{Tr}
\left(
\partial_x\hat P
\right)^2
+
\mathcal{V}_{\rm ani}[\hat P]\right).
\label{eq:CP3_density}
\end{align}
Here $\rho_s$ is the spin--valley stiffness arising from the dominant SU(4)-symmetric Coulomb exchange interaction, while $\mathcal{V}_{\rm ani}$ contains the weaker symmetry-breaking interactions.

\begin{table}[h]
\centering
\begin{tabular}{|lll|}
\hline
Term & Symmetry & Energy scale \\
\hline
\multicolumn{3}{c}{\textbf{(Dominant) Magnetic ordering energies}} \\
\hline
Long-range Coulomb interaction 
& $\mathrm{SU}(4)$ 
& $e^2/(\epsilon_r \ell_e) \sim 5$ meV \\
\hline
Single-particle band Hamiltonian 
& $\mathrm{SU}(2)_{K} \times \mathrm{SU}(2)_{K'} \times \mathrm{U}(1)_{\mathrm{v}}$  
& $\epsilon_F \sim 6$ meV \\
\hline
\multicolumn{3}{c}{\textbf{(Weak) Magnetic anisotropy energies}} \\
\hline
2-body lattice-scale interaction \cite{PhysRevB.111.125127}
& $\mathrm{SU}(2)_{\mathrm{s}} \times \mathrm{U}(1)_{\mathrm{v}}$  
& $g_\perp|n_e| \sim0.25$ meV \\
\hline
Spin–orbit coupling \cite{auerbach2025isospin}
& $\mathrm{U}(1)_{\mathrm{s}} \times \mathrm{U}(1)_{\mathrm{v}}$ 
& $\lambda_{\rm soc}\sim 0.1$~meV \\
\hline
\end{tabular}
\caption{
Hierarchy of approximate internal symmetries of the rhombohedral graphene Hamiltonian.
The leading magnetic energy scales arise from the competition between the long-range Coulomb interaction and the single-particle band Hamiltonian.
The weaker terms listed below introduce anisotropies that lift the $\mathrm{SU}(2)_{K} \times \mathrm{SU}(2)_{K'} \times \mathrm{U}(1)_{\mathrm{v}}$ symmetry.
The dominant energy scales are estimated for density–displacement field $n_e–U_D = (-0.9\times10^{11}\rm cm^{-2},25~ \rm meV)$ in bilayer graphene.
}\label{tab:interaction_hierarchy}
\end{table}
The hierarchy of interaction scales summarized in Table~\ref{tab:interaction_hierarchy} is a universal feature of interacting multilayer graphene systems, both twisted and untwisted, and provides the organizing principle for the continuum theory developed here. The dominant long-range Coulomb interaction establishes the spin--valley ferromagnetic order and determines the exchange stiffness, whereas lattice-scale interactions, intrinsic spin--orbit coupling (SOC), and intervalley Hund's exchange are parametrically weaker and primarily determine the local anisotropy of the order parameter. Guided by this hierarchy, we retain only the leading symmetry-allowed anisotropy terms,
\begin{align}
\mathcal{V}_{\rm ani}
=\frac{\mathcal K}{2}
\left\{
1-\left(\Tr[\hat P\hat{\tau}_z]\right)^2
\right\}
+
\frac{\lambda_{\rm soc}}{2}
\left\{
1-\Tr[\hat P\hat{s}_z\hat{\tau}_z]
\right\}
-u_\perp\,\mathbf S_K\cdot\mathbf S_{K'} .
\label{eq:CP3_anis}
\end{align}
Here
\begin{align}
\vec S_\tau
=
\mathrm{Tr}
\left(
\hat P
\hat{\mathcal P}_\tau
\hat{\vec s}
\right)
\end{align}
denotes the spin polarization in valley $\tau$, where $\hat{\mathcal P}_\tau=(1+\tau\hat\tau_z)/2$ is the valley projection operator.

The first term in Eq.~\eqref{eq:CP3_anis} is the energy cost of reducing the
Ising valley polarization from its uniform-ground-state value
$|\tau_z|=1$, where
$\tau_z\equiv\Tr[\hat P\hat{\tau}_z]$.
It therefore favors an Ising valley ferromagnet. The second term is the intrinsic Kane--Mele SOC, which locks the spin and valley quantization axes through the Ising coupling $s_z\tau_z$. The last term describes the intervalley Hund's interaction, favoring parallel alignment of the valley-resolved spin polarizations.

The continuum coefficients $(\rho_s,\mathcal{K},\lambda_{\rm soc})$ are determined directly from constrained microscopic Hartree--Fock calculations, providing a quantitative connection between the microscopic and continuum descriptions. The following subsection also establishes the correspondence between the microscopic intervalley Hund coupling $g_\perp$ and the continuum coupling $u_\perp$, thereby completing the parameterization of the $CP^3$ free-energy functional.

\subsection{Microscopic Parameterization}

Once the form of the $CP^3$ free-energy functional is established, its parameters are determined from the microscopic Hartree--Fock (HF) theory. Throughout this subsection, $E$ denotes the HF mean-field energy density, in contrast to the continuum free-energy functional $\mathcal{F}$ introduced in the previous subsection. All energy differences are measured relative to the
corresponding uniform ground state at the same carrier density.
Dividing the energy-density difference by $|n_e|$ gives the
energy difference per charge. The exchange stiffness $\rho_s$ and valley anisotropy coefficient $\mathcal{K}$ are determined independently from constrained microscopic calculations, while the continuum intervalley Hund coupling $u_\perp$ is related analytically to the microscopic interaction strength $g_\perp$. Intrinsic spin--orbit coupling strength $\lambda_{\rm soc}$ taken directly from experiment rather than extracted from the microscopic calculation.

The exchange stiffness originates from the dominant SU(4)-symmetric component of the long-range Coulomb exchange interaction. It is extracted from the energy cost of a long-wavelength spin spiral imposed on the uniform HF ground state. For sufficiently small spiral wave vector $q$, the increase in the HF energy density is quadratic,
\begin{align}
\frac{E(q)-E(0)}{|n_e|}
=
\frac{\rho_s}{2}q^2
+
\mathcal O(q^4),
\end{align}
from which $\rho_s$ is obtained by fitting the long-wavelength dispersion.

The valley-anisotropy coefficient $\mathcal K$ is determined
from the energy of spatially uniform Hartree--Fock states with
valley polarization 
$\tau_z\equiv\Tr[\hat P\hat{\tau}_z]$,
while keeping the carrier density fixed.  We subtract the
energy of the fully valley-polarized uniform ground state,
$\tau_z=1$, and plot the resulting energy difference per charge
against $1-\tau_z^2$.  The anisotropy energy then gives
\begin{align}
\frac{E(\tau_z)-E(\tau_z=1)}{|n_e|}
=
\frac{\mathcal K}{2}\left(1-\tau_z^2\right).
\end{align}

\begin{figure}[t]
    \centering
    \includegraphics[width=0.7\linewidth]{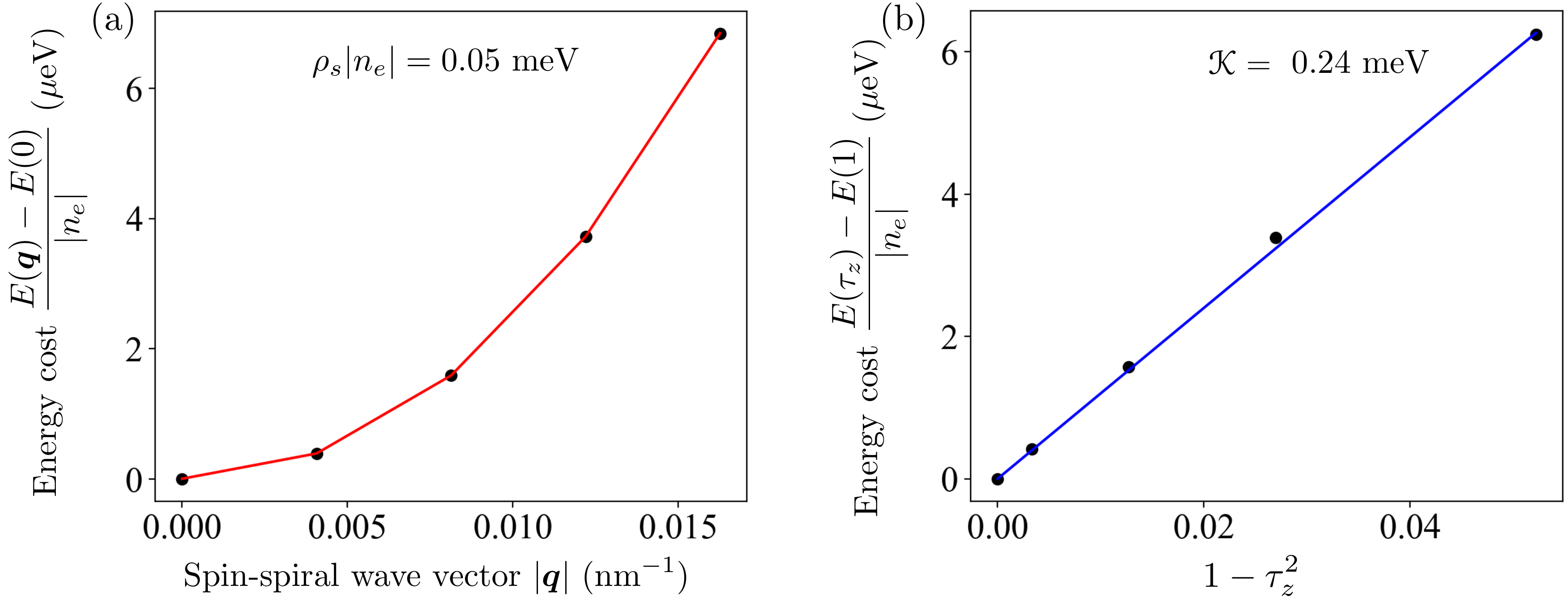}
    \caption{
    Microscopic extraction of the continuum parameters entering the $CP^3$ theory.
    (a) Hartree--Fock energy density of a long-wavelength spin spiral as a function of the spiral wave vector $q$. The quadratic fit yields the exchange stiffness $\rho_s$.
    (b) Uniform Hartree--Fock energy density as a function of $1-\tau_z^2$, where
$\tau_z=\Tr[\hat P\hat{\tau}_z]$ is valley polarization.
The straight line is a linear fit.  According to
$\Delta E/|n_e|=(\mathcal K/2)(1-\tau_z^2)$, the fitted slope
equals $\mathcal K/2$, and hence
$\mathcal K=2\times0.12~ \rm meV=0.24~ \rm meV$.
    }
    \label{fig:cp3_parameterization}
\end{figure}

The microscopic calculations used to extract $\rho_s$ and $\mathcal K$ are summarized in Fig.~\ref{fig:cp3_parameterization}. Panel (a) shows the quadratic dependence of the HF energy density on the spin-spiral wave vector, yielding the exchange stiffness $\rho_s={\rm 0.05~meV}/|n_e|\sim 50~{\rm meVnm^2}$. Panel (b) shows
$\bigl[E(\tau_z)-E(\tau_z=1)\bigr]/|n_e|$
as a function of $1-\tau_z^2$.  The data are linear, with slope
$\mathcal K/2$; therefore the valley-anisotropy energy is
$\mathcal K=2\times(\text{slope})$. The fitted slope is
$m=0.12\,{\rm meV}$, giving
$\mathcal K=0.24\,{\rm meV}$.

The intrinsic SOC strength is taken directly from experiment rather than extracted from the microscopic calculation. Throughout this work we use $\lambda_{\rm soc}=0.1$ meV, consistent with the experimentally estimated value\cite{auerbach2025isospin} for rhombohedral graphene.

The remaining continuum parameter, the intervalley Hund coupling $u_\perp$, is obtained by matching the microscopic intervalley interaction energy to the corresponding term in the $CP^3$ free-energy functional. To establish this correspondence, we evaluate the Hartree and Fock contributions to the microscopic mean-field energy for a uniform state and rewrite the result in terms of the continuum order parameters. Using Eqs.~(\ref{eq:Hund_H_uniform}) and (\ref{eq:Hund_F_uniform}), the Hartree and Fock contributions to the mean-field energy density are
\begin{align}
E^{\rm H,Hund}
&=
\frac{g_\perp}{2A^2}
\sum_{\tau,l,\sigma}
\sum_{\vec k,s}
\braket{c^\dagger_{\bar\tau sl\sigma}(\vec k)\,
c_{\tau sl\sigma}(\vec k)}
\sum_{\vec k',s'}
\braket{c^\dagger_{\tau s'l\sigma}(\vec k')\,
c_{\bar\tau s'l\sigma}(\vec k')},
\\
E^{\rm F,Hund}
&=
-\frac{g_\perp}{2A^2}
\sum_{\tau,l,\sigma}
\sum_{\vec k,s,s'}
\braket{c^\dagger_{\bar\tau sl\sigma}(\vec k)\,
c_{\bar\tau s'l\sigma}(\vec k)}
\sum_{\vec k'}
\braket{c^\dagger_{\tau s'l\sigma}(\vec k')\,
c_{\tau sl\sigma}(\vec k')}.
\label{eq:hund_selfenergy_mapping}
\end{align}

For the displacement fields considered in this work ($U_D\gg\epsilon_F$), the low-energy carriers are predominantly polarized onto a single layer--sublattice orbital. We therefore suppress the layer and sublattice indices $(l,\sigma)$ in the following discussion. We first introduce the valley-coherence components,
\begin{align}
\tau_x
&=
\frac{1}{N_e}
\sum_{\vec k,s,\tau}
\braket{
c^\dagger_{\bar\tau s}(\vec k)
c_{\tau s}(\vec k)
},
\\
\tau_y
&=
-\frac{i}{N_e}
\sum_{\vec k,s,\tau}
\tau\,
\braket{
c^\dagger_{\bar\tau s}(\vec k)
c_{\tau s}(\vec k)
},
\end{align}
together with the valley occupation and valley-resolved spin polarization,
\begin{align}
n_\tau
&=
\frac{1}{N_e}
\sum_{\vec k,s}
\braket{
c^\dagger_{\tau s}(\vec k)
c_{\tau s}(\vec k)
},
\\
\vec S_\tau
&=
\frac{1}{N_e}
\sum_{\vec k,s,s'}
\braket{
c^\dagger_{\tau s}(\vec k)
\hat{\vec s}_{ss'}
c_{\tau s'}(\vec k)
},
\end{align}
where $N_e=|n_e|A$ is the total number of charge carriers. The valley-resolved spin density matrix can then be written as
\begin{align}
\hat\rho_\tau
=
\frac{N_e}{2}
\left(
n_\tau\hat{\mathbb I}
+
\vec S_\tau\cdot\hat{\vec s}
\right).
\end{align}
Here $n_\tau$ denotes the fraction of carriers occupying valley $\tau$, satisfying $n_K+n_{K'}=1$. Because the valley-resolved spin polarization is normalized by the total carrier density, its magnitude satisfies $|\vec S_\tau|\le1$.

Expressed in terms of the valley coherence, the Hartree contribution becomes
\begin{align}
E^{\rm H,Hund}
=
\frac{g_\perp N_e^2}{4A^2}
\left(
\tau_x^2+\tau_y^2
\right).
\label{eq:hund_energy_cp3}
\end{align}
Similarly, the Fock contribution can be written as
\begin{align}
E^{\rm F,Hund}
&=
-\frac{g_\perp}{A^2}
\Tr_s
\left(
\hat\rho_K
\hat\rho_{K'}
\right)
\nonumber\\
&=
-\frac{g_\perp N_e^2}{2A^2}
\left(
n_Kn_{K'}
+
\vec S_K\cdot\vec S_{K'}
\right).
\label{eq:hund_fock_observables}
\end{align}
Using the definition of the valley polarization,
\begin{align}
\tau_z=n_K-n_{K'},
\end{align}
the valley occupations satisfy
\begin{align}
n_Kn_{K'}
=
\frac14
\left(
1-\tau_z^2
\right),
\end{align}
so that
\begin{align}
E^{\rm F,Hund}
=
-\frac{g_\perp N_e^2}{2A^2}
\left[
\frac14
\left(
1-\tau_z^2
\right)
+
\vec S_K\cdot\vec S_{K'}
\right].
\end{align}

Combining the Hartree and Fock contributions gives
\begin{align}
E^{\rm Hund}
=&
\frac{g_\perp N_e^2}{4A^2}\left[
\left(
\tau_x^2+\tau_y^2
\right)-
\frac{1}{2}
\left(
1-\tau_z^2
\right)\right]
-
\frac{g_\perp N_e^2}{2A^2}
\mathbf S_K\cdot\mathbf S_{K'}.
\label{eq:Hund_E1}
\end{align}
We separate the microscopic Hund contribution into a valley-only
part and a spin-exchange part. The valley-only contribution is treated
as a small correction to the valley-anisotropy energy obtained from the
microscopic Hartree--Fock theory and is included in the anisotropy
sector of the continuum model. To determine $u_\perp$, we match only
the spin-exchange contribution to the corresponding continuum Hund
term:
\begin{align}
\frac{E_{\rm spin}^{\rm F,Hund}}{|n_e|}
=
-\frac{g_\perp |n_e|}{2}
\mathbf S_K\cdot\mathbf S_{K'}.
\end{align}
Comparison with the corresponding term in
Eq.~(\ref{eq:CP3_anis}) then gives
\begin{align}
u_\perp
=
\frac{g_\perp |n_e|}{2}.
\end{align}

Together with the exchange stiffness $\rho_s$, the valley-anisotropy
coefficient $\mathcal K$, and the intrinsic SOC strength
$\lambda_{\rm soc}$ obtained from the microscopic Hartree--Fock
theory, this coefficient matching specifies the $CP^3$ energy
functional used in this work. The resulting parameters are summarized
in Table~\ref{tab:cp3_parameters}.
\begin{table}[b]
    \centering
    \begin{tabular}{|c|c|c|c|c|}
    \hline
    Parameter & $\rho_s|n_e|$ & $\mathcal{K}$ & $\lambda_{\rm soc}$ & $u_\perp$\\
    \hline
    Value (meV) & 0.05 & 0.24 & 0.10 & 0 -- 0.3\\
    \hline
    \end{tabular}
    \caption{
    Parameters entering the $CP^3$ continuum theory. The exchange stiffness $\rho_s$ and valley anisotropy coefficient $\mathcal{K}$ are extracted from microscopic HF calculations at $n$--$U_D=(-0.9\times10^{11}\,\mathrm{cm}^{-2},25~\mathrm{meV})$ in rBG. The SOC strength is taken from experiment\cite{auerbach2025isospin}, while the intervalley Hund coupling $u_\perp$ is treated as the tuning parameter.
    }
    \label{tab:cp3_parameters}
\end{table}

In the following subsection, we employ this parameterized continuum theory to investigate the reconstruction of magnetic domain walls as a function of the intervalley Hund coupling $u_\perp$.

\subsection{Domain-Wall Reconstruction in the $CP^3$ Theory}

\begin{figure}[h]
    \centering
    \includegraphics[width=\linewidth]{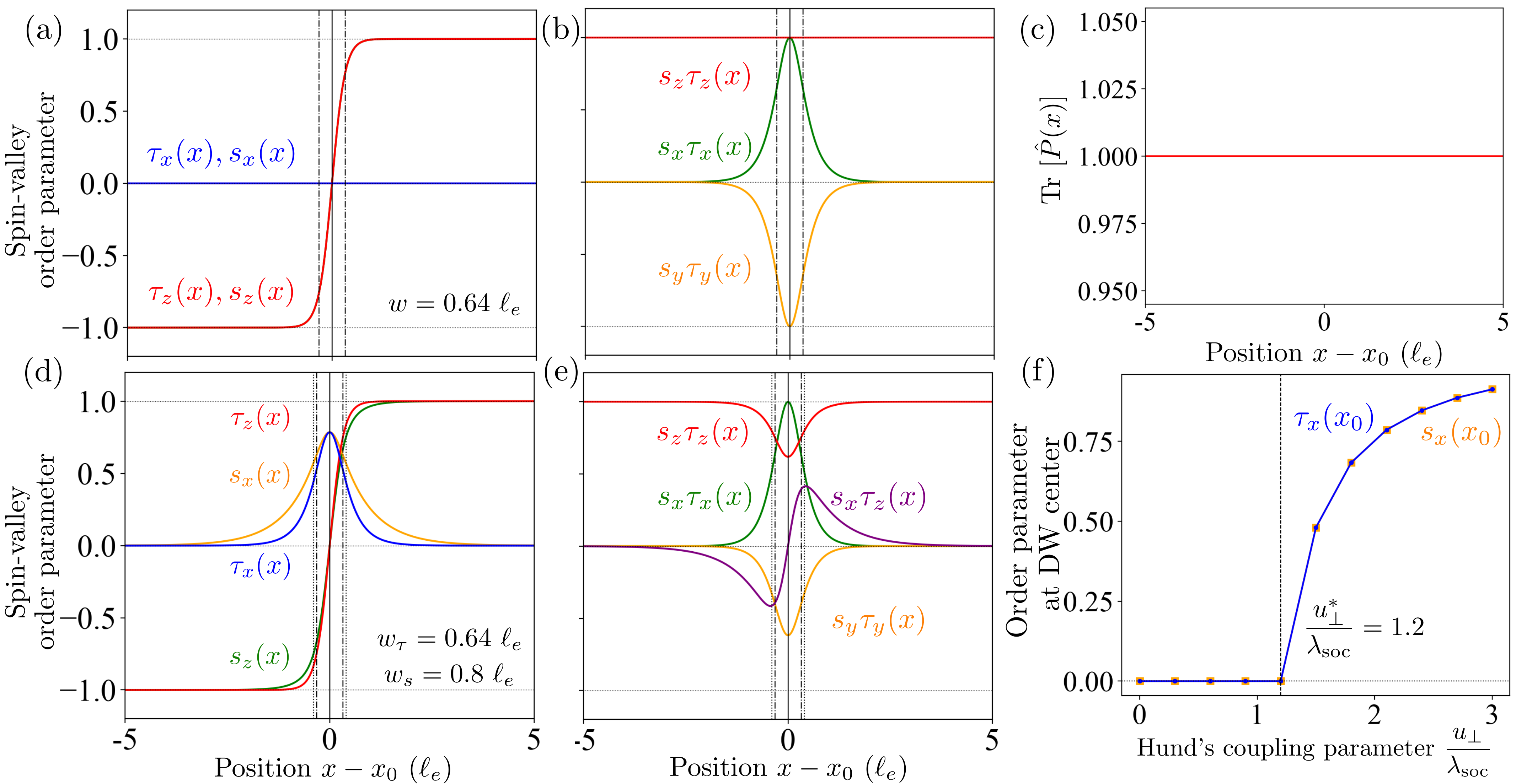}
    \caption{
$CP^3$ continuum domain-wall solutions obtained by minimizing the parameterized free-energy functional.
(a) Spatial profiles of the spin and valley order parameters for the two-component domain wall at $u_\perp=0$. The spin and valley remain locked, satisfying $s_z(x)=\tau_z(x)$ and $s_x(x)=\tau_x(x)=0$, resulting in a single domain-wall width $w$.
(b) Corresponding composite spin--valley order parameters. The Ising order parameter $s_z\tau_z(x)$ remains $1$ throughout the texture, while the transverse components vanish identically.
(c) Normalization of the $CP^3$ spin–valley projector, $\mathrm{Tr}[\hat P(x)]$. The continuum theory retains only the normalized internal spin–valley state and assumes a spatially uniform carrier density; charge redistribution such as that found microscopically in \ref{fig:cDWs}(c) is therefore excluded.
(d) Spatial profiles of the four-component domain wall for $u_\perp/\lambda_{\rm soc}=2.1$. Finite transverse spin polarization and intervalley coherence develop inside the wall, leading to distinct spin and valley domain-wall widths, $w_s>w_\tau$.
(e) Corresponding composite spin--valley order parameters for the four-component domain wall.
(f) Evolution of the domain-wall-center order parameters
$s_x(x_0)$ and $\tau_x(x_0)$ as functions of the dimensionless
Hund coupling $u_\perp/\lambda_{\rm soc}$.
Above the critical coupling
$u_\perp^*/\lambda_{\rm soc}\approx1.2$,
the transverse order parameters develop continuously from zero,
signaling a second-order transition from the two-component to the
four-component domain wall. Using the relation
$u_\perp=g_\perp|n_e|/2$,
the critical coupling corresponds to
$g_{\perp,\rm CP^3}^*|n_e|/\lambda_{\rm soc}=2.4$.
}
    \label{fig:cp3_DWs}
\end{figure}

We now minimize $CP^3$ free-energy functional (Eq.~\eqref{eq:CP3_density}) subject to the same $K\uparrow$ and $K'\downarrow$ boundary conditions imposed in the microscopic Hartree--Fock calculations. Since this functional is invariant under
$U(1)_s\times U(1)_v$ rotations about the spin and valley
$z$ axes, we choose a symmetry-equivalent representative with
$s_y(x)=\tau_y(x)=0$, so that both polarizations lie in their
respective $x$--$z$ planes. The normalized spinor is initialized as a smooth interpolation between the two bulk spin--valley polarized quarter-metal states and is iteratively relaxed until the free energy converges. Figure~\ref{fig:cp3_DWs} summarizes the resulting continuum domain-wall solutions.

For weak intervalley Hund coupling, the minimizing spinor remains confined to the $\{K\uparrow,K'\downarrow\}$ subspace, giving rise to the two-component domain wall shown in Figs.~\ref{fig:cp3_DWs}(a)--(c). The spin and valley order parameters remain locked throughout the texture, satisfying $s_z(x)=\tau_z(x)$ and $s_x(x)=\tau_x(x)=0$. Consequently, the Ising order parameter $s_z\tau_z(x)$ remains unity across the entire domain wall, while the intervalley coherence vanishes identically. As expected within the continuum theory, the local charge density remains uniform throughout the texture, as shown in Fig.~\ref{fig:cp3_DWs}(c). The spin and valley profiles are therefore characterized by the same domain-wall width.

Upon increasing the intervalley Hund coupling, the energy gained by aligning the valley-resolved spin polarizations competes with the spin--valley locking induced by intrinsic SOC. Above the critical coupling
$u_\perp^*/\lambda_{\rm soc}\approx1.2$,
the energy minimum continuously expands into the full $CP^3$ manifold, producing the four-component domain wall shown in Figs.~\ref{fig:cp3_DWs}(d) and (e). In this regime, finite transverse spin polarization $s_x(x)$ and intervalley coherence $\tau_x(x)$ develop within the domain wall, accompanied by the composite order parameters $s_x\tau_x(x)$, $s_y\tau_y(x)$, and $s_x\tau_z(x)$. The spin and valley order parameters are no longer locked, resulting in distinct domain-wall widths, with $w_s>w_\tau$.

The continuous nature of this reconstruction is illustrated in Fig.~\ref{fig:cp3_DWs}(f), which shows the evolution of the domain-wall-center order parameter $s_x(x_0)$ and $\tau_x(x_0)$ as a function of the dimensionless Hund coupling $u_\perp/\lambda_{\rm soc}$. Below the critical coupling, the transverse order parameter vanishes and the stable solution is the two-component domain wall. Above $u_\perp^*/\lambda_{\rm soc}$, $s_x(x_0)$ and $\tau_x(x_0)$ develop continuously from zero, signaling a second-order transition into the four-component domain wall. Overall, the parameterized $CP^3$ theory successfully reproduces the microscopic Hartree--Fock results, including the two distinct classes of domain walls, the spin--valley unlocking inside the four-component domain wall ($w_s>w_\tau$), and the continuous Hund-driven transition between them. These results demonstrate that the domain-wall reconstruction is a generic consequence of the competition between spin--orbit coupling and intervalley Hund exchange, rather than a feature specific to the microscopic details of rhombohedral graphene.

Although the parameterized $CP^3$ theory successfully captures the
qualitative domain-wall reconstruction obtained in the microscopic
Hartree--Fock calculations, it is not expected to reproduce every
quantitative feature of the microscopic solution. The continuum theory
retains only the local spin--valley order parameter through a rank-one
projector satisfying $\hat P^2=\hat P$, while assuming a uniform local
carrier density. By contrast, the microscopic Hartree--Fock theory
explicitly incorporates the layer, sublattice, and momentum degrees of
freedom. After tracing over the layer--sublattice sector, the resulting
local spin--valley density matrix becomes mixed near the domain-wall
center, as quantified by the finite value of
$\operatorname{Tr}(\hat P_{\rm HF}-\hat P_{\rm HF}^2)$ shown in
Fig.~\ref{fig:dw_fig_3} of the main text.
The self-consistent Hartree--Fock solution also exhibits a weak
redistribution of the local charge density
[Fig.~\ref{fig:cDWs}(c) of the main text], which contributes a finite
Hartree energy to the domain-wall line energy. The $CP^3$ theory instead
assumes a frozen local carrier density and therefore contains no
corresponding Hartree contribution. These additional microscopic
degrees of freedom lead to quantitative differences between the two
descriptions. In particular, the domain-wall widths predicted by the
continuum theory differ from the microscopic Hartree--Fock results by
approximately a factor of two.
Furthermore, using the microscopic relation
$u_\perp=g_\perp|n_e|/2$, the continuum critical coupling
$u_\perp^*/\lambda_{\rm soc}\approx1.2$ corresponds to
\begin{align}
\frac{g_{\perp,\rm CP^3}^*|n_e|}{\lambda_{\rm soc}}
\approx 2.4,
\end{align}
whereas the microscopic Hartree--Fock calculation gives
$g_{\perp,\rm HF}^*|n_e|/\lambda_{\rm soc}\approx11$. Thus, the
continuum estimate of the critical microscopic coupling is smaller by
a factor of approximately $11/2.4\approx4.6$. Moreover, the rank-one $CP^3$ pure-state constraint, together with the
symmetries of the wall-center solution, yields equal transverse spin
polarization and intervalley coherence,
$|s_x(x_0)|=|\tau_x(x_0)|$. The microscopic Hartree--Fock solution
generally yields different magnitudes for these order parameters
because its local reduced spin--valley density matrix is no longer
pure.

Nevertheless, the two approaches are in remarkable qualitative agreement. Both predict two distinct classes of spin--valley domain walls connected by a continuous Hund-driven transition, together with the characteristic spin--valley unlocking inside the four-component domain wall. This agreement demonstrates that the reconstruction is governed primarily by the competition between the exchange stiffness, spin--orbit-induced spin--valley locking, and intervalley Hund exchange, rather than by microscopic details of the electronic band structure. The resulting $CP^3$ theory therefore identifies the minimal collective ingredients required to describe interaction-driven spin--valley domain-wall reconstruction in graphene ferromagnets.

\end{document}